%% file: main.tex
\documentclass[letterpaper,twocolumn,10pt]{article}
\usepackage{usenix2019_v3}

\input{commands}
\begin{document}
%-------------------------------------------------------------------------------

\setlength{\abovedisplayskip}{5pt}
\setlength{\belowdisplayskip}{5pt}
\setlength{\abovedisplayshortskip}{3pt}
\setlength{\belowdisplayshortskip}{3pt}

%don't want date printed
\date{}

% \title{\Large \bf \toolnospace: A Unified Agentic Framework for Multi-Task Network Traffic Analysis}
\title{\Large \bf \toolnospace: Multi-Task Agentic Network Traffic Analysis Made Practical}

\author{
{\rm Hao Fu}\\
Virginia Tech\\
haofu26@vt.edu
\and
{\rm Dawn Song}\\
UC Berkeley\\
dawnsong@berkeley.edu
\and
{\rm Peng Gao}\\
Virginia Tech\\
penggao@vt.edu
} % end author

\maketitle

\input{sections/abstract}

\input{sections/introduction}

\input{sections/background}

\input{sections/system_design}
\input{sections/evalution}
\input{sections/discussion}

\input{sections/conclusion}
\cleardoublepage

\bibliographystyle{plain}
\bibliography{references}

\appendix
\input{sections/app_toolkit}

\input{sections/app_prompts}

\end{document}

%% file: commands.tex
\usepackage{tikz}
\usepackage{amsmath}

\usepackage{filecontents}
\usepackage{tablefootnote}

\usepackage{microtype}
\usepackage{algorithm}
\usepackage[spaceRequire=false]{algpseudocodex}
\usepackage[utf8]{inputenc} 
\usepackage{listings}
\usepackage{xspace}
\usepackage{subcaption}
\usepackage{soul}
\usepackage{enumitem}
\usepackage{booktabs}
\usepackage{multirow}
\usepackage[most]{tcolorbox}
\usepackage[subtle]{savetrees}
\usepackage{makecell}
\usepackage{balance}
\usepackage{adjustbox}

\usepackage[table]{xcolor}
\definecolor{challengebg}{HTML}{DCE6F1}
\newcommand{\challenge}[1]{{\sethlcolor{challengebg}\hl{#1}}}

\usepackage[labelfont=bf,skip=5pt,figurename=Fig.]{caption}
\usepackage[capitalize,nameinlink]{cleveref}
\hypersetup{%
	bookmarksnumbered, bookmarksopen=true, bookmarksopenlevel=1,%
}
\crefname{figure}{Fig.}{Figs.}
\crefname{listing}{Listing}{Listings}
\crefname{section}{\S}{Sections}
\crefname{table}{Table}{Tables}
\crefname{algorithm}{Algorithm}{Algorithms}
\crefname{line}{Line}{Lines}
\crefname{equation}{Eq.}{Eqs.}
\crefname{rq}{RQ}{RQs}
\crefname{appendix}{Appendix}{Appendices}

\usepackage{tikz}
\usepackage{amsmath}
\usepackage{amssymb}
\usepackage{booktabs}
\usepackage{tabularx}
\usepackage{array}
\usepackage{graphicx}
\usepackage{subcaption}
\usepackage{flafter}

\newtcolorbox{takeaway}{enhanced, breakable,
  colback=black!8, coltext=black,
  boxrule=0.4pt, colframe=black!25, arc=3pt, left=4pt, right=4pt, top=2pt, bottom=2pt,
  fontupper=\small, before skip=1.5ex, after skip=1.5ex}
  
\graphicspath{{pic/}}
\usepackage{listings}
\lstdefinestyle{flowprompt}{
  basicstyle=\ttfamily\fontsize{6.5pt}{8pt}\selectfont,
  breaklines=true, breakatwhitespace=false, keepspaces=true,
  columns=flexible, showstringspaces=false,
  frame=single, framerule=0.4pt, rulecolor=\color{black!25},
  backgroundcolor=\color{black!4},
  xleftmargin=4pt, xrightmargin=4pt, aboveskip=4pt, belowskip=6pt,
}

\newcounter{flowalgo}

\renewcommand{\paragraph}[1]{\smallskip\noindent\textbf{#1.}}

\newcommand{\tool}{NetAgent\xspace}
\newcommand{\toolnospace}{NetAgent}

\newif\ifshowcomment
\showcommenttrue
\ifshowcomment
    \newcommand{\pgao}[1] {{\footnotesize\color{red}[Peng: #1]}}
    \newcommand{\fh}[1]{{\footnotesize\color{purple}[Fu: #1]}}
\else
    \newcommand{\pgao}[1]{}
    \newcommand{\dawn}[1]{}
    \newcommand{\fh}[1]{}
    \newcommand{\dawn}[1]{{\footnotesize\color{blue}[Dawn: #1]}}
\fi

\newcommand{\eat}[1]{}

\usepackage{pifont}
\usepackage{graphicx}

\usepackage[dvipsnames,table]{xcolor}
\usepackage{colortbl}
\usepackage[most]{tcolorbox}
\usepackage{listings}

\definecolor{tableheader}{RGB}{182,216,210}
\definecolor{tablerow}{RGB}{220,239,235}
\definecolor{tablebest}{RGB}{195,225,219}

\definecolor{pastelblue}{RGB}{185,195,235}
\definecolor{pastelbluebg}{RGB}{232,236,250}
\definecolor{pastelgreen}{RGB}{175,220,200}
\definecolor{pastelgreenbg}{RGB}{228,245,236}
\definecolor{pastelrose}{RGB}{230,185,185}
\definecolor{pastelrosebg}{RGB}{248,232,232}
\definecolor{promptgray}{RGB}{245,245,248}
\definecolor{promptframe}{RGB}{200,200,210}

\newtcolorbox{systempromptbox}[1][]{%
  enhanced, breakable,
  colback=pastelbluebg, colframe=pastelblue, coltitle=black,
  fonttitle=\bfseries\small, title={#1},
  boxrule=0.6pt, arc=2pt,
  left=4pt, right=4pt, top=2pt, bottom=2pt,
  toptitle=2pt, bottomtitle=2pt,
  attach boxed title to top left={yshift=-2mm, xshift=4mm},
  boxed title style={colback=pastelblue, colframe=pastelblue, arc=1.5pt, boxrule=0pt},
}
\newtcolorbox{userpromptbox}[1][]{%
  enhanced, breakable,
  colback=pastelgreenbg, colframe=pastelgreen, coltitle=black,
  fonttitle=\bfseries\small, title={#1},
  boxrule=0.6pt, arc=2pt,
  left=4pt, right=4pt, top=2pt, bottom=2pt,
  toptitle=2pt, bottomtitle=2pt,
  attach boxed title to top left={yshift=-2mm, xshift=4mm},
  boxed title style={colback=pastelgreen, colframe=pastelgreen, arc=1.5pt, boxrule=0pt},
}
\newtcolorbox{retrypromptbox}[1][]{%
  enhanced, breakable,
  colback=pastelrosebg, colframe=pastelrose, coltitle=black,
  fonttitle=\bfseries\small, title={#1},
  boxrule=0.6pt, arc=2pt,
  left=4pt, right=4pt, top=2pt, bottom=2pt,
  toptitle=2pt, bottomtitle=2pt,
  attach boxed title to top left={yshift=-2mm, xshift=4mm},
  boxed title style={colback=pastelrose, colframe=pastelrose, arc=1.5pt, boxrule=0pt},
}

\lstdefinestyle{promptstyle}{
  basicstyle=\ttfamily\scriptsize,
  breaklines=true, breakatwhitespace=false,
  columns=fullflexible, keepspaces=true,
  frame=none, xleftmargin=0pt,
  aboveskip=0pt, belowskip=0pt,
}

%% file: sections/abstract.tex
\begin{abstract}

Network traffic analysis is central to network security, spanning tasks from intrusion detection to encrypted traffic classification. Existing approaches either train task-specific models that generalize poorly or rely on costly traffic foundation models that still struggle under distribution shift. We present \tool, the first agentic framework for multi-task traffic analysis. Through a carefully designed agent loop, \tool supports complex task understanding, on-the-fly decomposition and orchestration, dynamic replanning, and long-horizon analysis, without task-specific training. It introduces five key designs: (1) \emph{knowledge-augmented workflow planning} that maps attack knowledge to traffic features to bridge the semantic gap; (2) a comprehensive \emph{tool action space} with 150+ verified tools extracted from 50+ published systems; (3) a unified \emph{code execution space} for flexible action composition; (4) a \emph{three-tier memory} for long-term knowledge consolidation; and (5) sandboxing and runtime repair for \emph{reliable execution}.

Across 9 major benchmarks, \tool outperforms all baselines (23 single-task and 5 multi-task) on nearly all tasks and generalizes substantially better to unseen traffic distribution (90.04\% F1 vs. 2.74\% and 3.04\% for the best single-task and multi-task baselines) and under realistic background shift (4.85-point F1 drop vs. 74.88-point and 74.80-point drop for the best single-task and multi-task baselines). These results reveal that existing methods owe much of their reported success to overfitting dataset-specific patterns and degrade sharply in realistic network environments, while \tool's agentic design remains accurate, generalizable, and robust.

\end{abstract}

%% file: sections/introduction.tex
%-------------------------------------------------------------------------------
\section{Introduction}
%-------------------------------------------------------------------------------

Network traffic analysis examines network communication data across diverse tasks such as intrusion detection~\cite{hypervision}, malicious traffic detection~\cite{pointcloud}, and encrypted traffic classification~\cite{rosetta}. A security operations center (SOC) often handles many of these tasks and synthesizes their results into conclusions. Existing methods typically train a separate machine learning model per task, using tree-based classifiers over flow statistics~\cite{iotdeviceid}, Graph Neural Networks (GNNs) over flow interactions~\cite{pointcloud}, or transformers over packet sequences~\cite{etbert}. This per-task paradigm limits knowledge transfer and composability across tasks, and every new task requires fresh annotation and retraining.

Two lines of work aim to unify traffic analysis tasks. nPrint~\cite{nprint} automates per-task model training with AutoML, but still produces isolated models and is limited to shallow classifiers without deep traffic understanding. TrafficFormer~\cite{trafficformer} pre-trains a large transformer and fine-tunes it for downstream tasks, but pre-training is costly, fine-tuning requires substantial task-specific labels, and the model remains vulnerable to distribution shift. Our experiments (Section~\ref{sec:eval_rq3}) show that TrafficFormer suffers an F1 drop of over 61.2 points under distribution shift. Moreover, such monolithic models map inputs to labels in a single forward pass and cannot decompose complex analyses into coordinated subtasks.

We argue that traffic analysis calls for a \emph{new paradigm} that delivers complex task decomposition, orchestration, and generalization. Consider auditing an enterprise network. The analyst may first classify encrypted traffic by application to filter out known applications, then run tunnel detection on the remaining unattributed flows to identify covert channels such as DNS tunnels, and finally analyze all traffic from hosts using those channels for malicious activity to determine whether they are compromised. These tasks build on one another and require coordination as an integrated workflow, which no individual model can provide.

Large language models (LLMs) are a natural foundation for this paradigm: they bring broad attack knowledge from pre-training, can interpret analyst intent, and reason over data without task-specific training. However, LLMs alone lack awareness of the live network environment and are limited in complex task planning and execution. \emph{LLM-based agents} address these limits by placing the LLM with its environment: an agent decomposes analyst queries, invokes diverse tools for processing, and synthesizes multi-task results through LLM reasoning. Such agents have shown strong performance in domains like software engineering~\cite{metagpt,autogen} and biomedical research~\cite{biomni}, yet no unified agent for multi-task traffic analysis has emerged.

\paragraph{Contributions} We present \textbf{\tool}, the first \textbf{unified agentic framework for multi-task network traffic analysis}. \tool structures traffic analysis into functional layers, from environmental perception to feature extraction, feature construction, and traffic classification, with uniform tool interfaces that enable flexible tool composition across layers. A carefully designed agent loop coordinates three agents, a planner, executor, and decision maker, to decompose an analyst query into orchestrated workflows, execute them, and decide whether to conclude or replan. Our \emph{key insight} is that diverse traffic analysis tasks share reusable operations, such as packet parsing, feature extraction, traffic aggregation, and flow classification. By exposing these operations as atomic tools, \tool enables the agents to dynamically compose them into workflows for solving different traffic tasks.

\tool addresses \challenge{key challenges} specific to the traffic domain and makes the following technical contributions:

\begin{enumerate}[label={\textbf{(\arabic*)}}, leftmargin=*, itemsep=1pt, topsep=1pt]

    \item \challenge{An agent is only as capable as its action space, the set of tools through which it perceives and acts, yet no such space exists for traffic analysis.} We thus construct a comprehensive \emph{tool action space} with an LLM-driven discovery pipeline that mines reusable operations from 50+ published traffic analysis systems, yielding a library of 150+ verified tools, covering critical capabilities across diverse traffic analysis tasks. 
    %(Table~\ref{tab:tools} in Appendix~\ref{app:results}).

    \item \challenge{In the multi-task setting with diverse attack patterns, the agent must bridge the semantic gap between high-level attack knowledge and low-level traffic features.} We introduce \emph{knowledge-augmented workflow planning}. Traffic is normalized into a unified multi-view feature matrix, and an \emph{attack knowledge base} expresses attack patterns as conditions over these features. For example, a long-lived connection with small packets and large inter-arrival times may indicate cryptomining. The planner retrieves relevant attack knowledge to augment workflow planning.

	\item \challenge{Typical agents emit actions as text or JSON in a predefined format, which limits their ability to dynamically compose multiple tools and revise prior actions.} \tool uses \emph{code as the composition medium}. The executor maintains a persistent code space in which all tools are callable functions and each action is emitted as a code snippet that flexibly composes tools and passes results. Execution self-corrects using interpreter feedback.
    
    \item \challenge{A long-horizon analysis must recover quickly from failures, as a single traffic data read error can cascade through subsequent steps and fail the entire analysis.} The executor uses sandboxed execution, static error checks, and dynamic repair to re-execute only the minimal set of steps affected according to the dataflow dependencies.
    
    \item \challenge{A long-horizon analysis must retain and reuse knowledge across tasks under continuous network drift.} \tool maintains a \emph{three-tier memory} that consolidates raw analysis conclusions into recurring behavior patterns and then into long-term knowledge. This layered consolidation compacts memory, enables knowledge reuse, and resolves drift-induced conflicts for history-aware decision making.    

\end{enumerate}

We evaluate \tool's performance, generalizability, and robustness on 9 widely used benchmarks spanning diverse traffic analysis tasks (Table~\ref{tab:datasets}). We systematically compare \tool against 28 baselines: 23 single-task methods, including 7 classical ML and 16 SOTA deep learning methods (e.g., GNNs~\cite{tfegnn,pointcloud}, transformers~\cite{etbert,multitabwf}, autoencoders~\cite{rapier,tfegnn}), and 5 multi-task methods, including nPrint~\cite{nprint}, TrafficFormer~\cite{trafficformer}, ET-BERT~\cite{etbert}, TrafficLLM~\cite{trafficllm}, and a ReAct agent~\cite{yao2022react} equipped with our tool library.
The results show that:
\begin{enumerate}[label={\textbf{(\arabic*)}}, leftmargin=*, itemsep=1pt, topsep=1pt]

    \item \textbf{\tool delivers consistently strong performance across diverse traffic analysis tasks.} It achieves the best performance on 8 of 9 tasks, outperforming all single-task and multi-task baselines with F1 above 98.63\%. On VPN traffic classification, \tool achieves 92.42\% F1, outperforming all single-task methods and all multi-task methods except ET-BERT (94.54\% F1); however, \tool significantly exceeds ET-BERT by an average of 35.51 F1 points across the other 8 tasks.

    \item \textbf{\tool generalizes substantially better to unseen traffic distribution.} When baselines are trained on UNSW-NB15 and tested on the unseen CIC-IDS2017 dataset for web attack detection, the best single-task and multi-task baselines achieve only 2.74\% and 3.04\% F1, while \tool achieves 90.04\%. 
    
    \item  \textbf{\tool remains robust under realistic background distribution shift.} After injecting 5$\times$ real-world MAWI/WIDE backbone traffic into the web attack detection test set, the best single-task and multi-task baselines lose 74.88 and 74.80 F1 points, while \tool drops by only 4.85 points. 
    
\end{enumerate}

These results demonstrate that prior methods often overfit dataset-specific patterns and degrade sharply in realistic network environments with background traffic noise and distribution shift, whereas \tool's agentic design remains accurate, generalizable, and robust. Beyond these gains, the same design covers a broad range of traffic analysis tasks in one unified framework without task-specific training.

%% file: sections/background.tex
%-------------------------------------------------------------------------------
\section{Background and Challenges}
%-------------------------------------------------------------------------------

\subsection{Network Traffic Analysis}
\label{subsec:tasks}

Understanding network traffic is fundamental to network defense because it carries much of the evidence needed to distinguish benign activity from attacks. Network traffic analysis examines this data to characterize traffic and identify threats. It operates on \emph{heterogeneous inputs}, from raw PCAP captures and NetFlow records to CSV or TXT feature tables exported by upstream collectors. Across these inputs, security teams perform a \emph{diverse set of analysis tasks}:

\begin{itemize}[leftmargin=*, itemsep=1pt, topsep=1pt]

\item
\textbf{Intrusion and malicious traffic detection} identify attacks within benign traffic, such as scanning~\cite{ruling}, brute force~\cite{ruling}, web attacks~\cite{ruling}, malware command-and-control~\cite{hypervision,pointcloud}, botnets~\cite{davanian2024c2miner}, cryptomining~\cite{mineshark}, and DDoS~\cite{lemon}.

\item
\textbf{Encrypted traffic classification} attributes flows to applications~\cite{rosetta} and services~\cite{tfegnn}, and identifies traffic carried through VPNs and proxies~\cite{proxyfp,crossrtt}, Tor and other tunnels~\cite{gfwencrypted}, and covert channels such as malicious DNS-over-HTTPS~\cite{gfwencrypted}.

\item
\textbf{Fingerprinting} infers entities and content behind flows. Website fingerprinting identifies websites~\cite{multitabwf}, individual pages~\cite{oscar}, and streamed videos~\cite{endangered}. Device fingerprinting identifies operating systems and hardware through probing or passive observation~\cite{nprint}, such as IoT devices~\cite{iotdeviceid,wifinger}.

\item
\textbf{Flow correlation} links traffic observed at different vantage points, for example tracing a flow across Tor~\cite{swallow}.

% \item While the tasks above operate on the data plane,  \textbf{routing anomaly detection} instead works at the control plane, identifying hijacks and leaks in BGP announcements~\cite{semanticsaware}. 

\end{itemize}

Beyond security, traffic data support other networking tasks, including synthetic traffic generation~\cite{trafficllm}, traffic forecasting to predict volumes and patterns~\cite{gui2024redte}, and traffic engineering for optimizing routing and resource allocation~\cite{aether}.

%%%%%%%%%%%%%%%%%%%%%%%%%%%%%%%%%%%
\subsection{Traffic Analysis Methods}

\paragraph{Rule and statistics-based methods} Early systems rely on expert-written rules and statistical heuristics. Port-based methods map flows to applications using registered ports~\cite{dreger2006dynamic}. Deep packet inspection systems such as Snort~\cite{roesch1999snort} and Suricata~\cite{suricata} match payloads against signatures. These methods are fast, but attackers can evade fixed rules by shifting ports or mutating payloads. Maintaining effective rule sets without overwhelming analysts with excessive false alerts remains an open problem~\cite{verissimo2007intrusion}.

\paragraph{Shallow machine learning} Later methods move beyond fixed rules by combining hand-crafted traffic features, such as packet lengths and inter-arrival times, with shallow ML classifiers such as decision trees and random forests. Examples include intrusion detection over flow statistics~\cite{jafri2024leo}, ML-based device fingerprinting~\cite{iotdeviceid,wifinger,nprint}, and proxy detection using handshake and timing features~\cite{proxyfp,crossrtt}. Their effectiveness, however, is limited by hand-crafted features, which need to be redesigned for new tasks and lose discriminative power as applications and attacks evolve under distribution shift.

\paragraph{Deep learning} More recent methods learn representations directly from raw bytes and packet sequences, capturing hidden patterns and achieving SOTA performance across many tasks. Examples include Graph Neural Networks (GNNs) that model structural relations among flows for malicious traffic detection and encrypted traffic classification~\cite{hypervision,tfegnn,pointcloud}, and sequence models that capture packet-level temporal patterns for DDoS detection and website fingerprinting~\cite{rosetta,multitabwf,oscar,swallow}. However, these methods still train a separate model per task and require large labeled datasets, making adaptation across multiple or new tasks costly.

\paragraph{Large pre-trained models} Another line of work pre-trains a shared model for multiple tasks. ET-BERT~\cite{etbert}, TrafficFormer~\cite{trafficformer}, and MM4flow~\cite{mm4flow} pre-train transformers on large traffic corpora with self-supervised objectives and fine-tune them for each downstream task. Instead of pre-training from scratch, a recent work, TrafficLLM~\cite{trafficllm} (unpublished), adapts a general-purpose LLM with traffic-specific tokenization and fine-tuning. However, large model pre-training is costly, fine-tuning still requires substantial task-specific labels, and monolithic models map inputs to labels in a single pass and cannot decompose complex analyses into coordinated steps. As our experiments show (Section~\ref{sec:eval_rq3}), both TrafficFormer and TrafficLLM suffer substantial generalization degradation under network drift.

%%%%%%%%%%%%%%%%%%%%%%%%%%%%%%%%%%%

\subsection{LLM Agents}

Large language models (LLMs) bring broad attack knowledge and strong reasoning capabilities. LLM-based agents further extend these capabilities by placing the LLM in an environment loop that plans subtasks, invokes tools, observes feedback, and revises its plan~\cite{yao2022react}, enabling multi-step analyses beyond a single model pass. Such agents have shown strong results across many domains including software engineering~\cite{metagpt,autogen}, biomedicine~\cite{biomni}, and chemical research~\cite{m2024augmenting}.

Agents have also emerged in security. PentestGPT~\cite{pentestgpt} assists human-guided penetration testing. PentestAgent~\cite{pentestagent} automates reconnaissance and exploitation. Chimera~\cite{chimera} simulates insider threats with multiple agents. The CAGE challenges~\cite{cage} benchmark reinforcement learning agents that defend simulated networks by acting on abstracted environment states rather than real traffic. In traffic analysis, agents have only begun to appear. IDS-Agent~\cite{idsagent} applies an iterative loop to IoT intrusion detection, but the loop in effect only aggregates predictions from a few shallow classifiers and targets IoT traffic alone. To date, no agent exists that supports multi-task traffic analysis across diverse task domains, heterogeneous inputs, and long-horizon complex analyses.

%%%%%%%%%%%%%%%%%%%%%%%%%%%%%%%%%%%

\input{sections/goal}

%% file: sections/goal.tex
\subsection{Goals and Challenges}
\label{subsec:challenges}

\begin{figure}[t]
\centering
\includegraphics[width=.85\columnwidth]{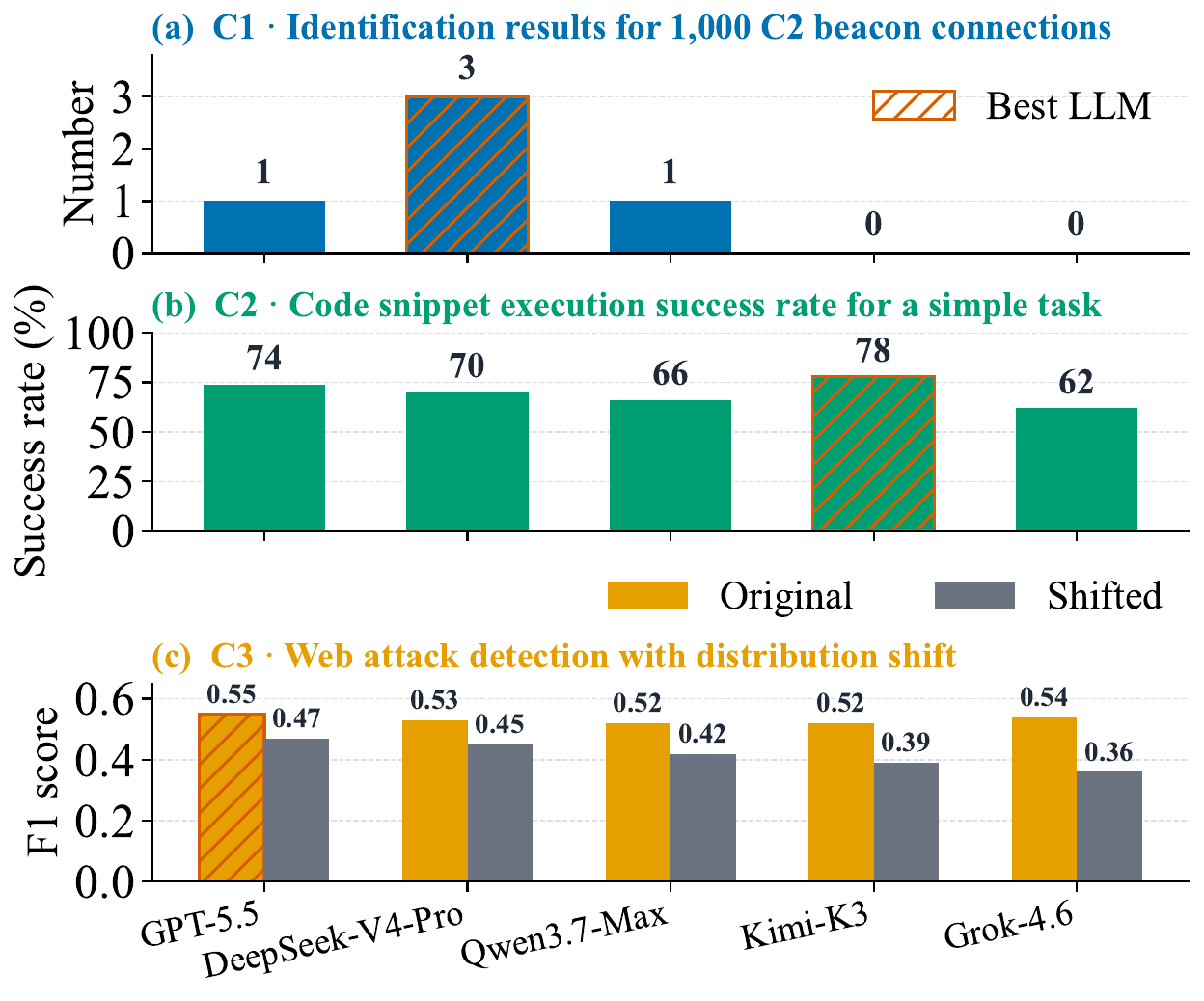}
\caption{Empirical validation of three challenges in directly applying LLMs to traffic analysis: attack misidentification and hallucination, low code execution success, and severe performance degradation under network drift.}
\label{fig:challenge_measurement}
\vspace{-1ex}
\end{figure}

We consider a network operator or analyst who needs to understand traffic crossing a vantage point, such as an enterprise gateway or SOC monitoring tap. Existing systems expose a fixed interface: they consume traffic and emit labels, offering little support for inspecting evidence behind a decision or pursuing follow-up questions. 
We instead target an \emph{interactive paradigm}, where the analyst queries captured traffic in natural language, receives a conclusion, and follows up with further questions. For example, the analyst may first ask which applications generate encrypted flows and, finding that some flows match no known application, follow up by asking whether they hide command-and-control traffic. Such analysis spans the heterogeneous inputs and diverse tasks in Section~\ref{subsec:tasks}. Applying LLMs or agents to this domain, however, introduces \textbf{three key challenges:}

\paragraph{C1. Semantic gap} LLMs encode high-level attack knowledge in natural language, while network captures expose only low-level packet bytes and traffic statistics. The \emph{semantic gap} lies in mapping attack concepts to the concrete traffic features that reveal them. For example, an LLM may know that malware periodically beacons to a command-and-control (C2) server, but detecting this behavior requires signals such as inter-arrival times and flow sizes across thousands of packets to identify connections with regular timing and consistently small payloads. Without this mapping, the LLM cannot reliably recognize beaconing in raw traffic.
We demonstrated this using Tinba banking trojan captures from USTC-TFC2016~\cite{ustctfc}. As Figure~\ref{fig:challenge_measurement}(a) shows, among 1,000 C2 beacon connections to DGA domains, five major LLMs identified at most three and hallucinated attacks absent from the capture.
%, such as abnormal TCP session termination and incomplete network discovery.

\paragraph{C2. Fragmented action space} Multi-task traffic analysis requires diverse external capabilities, such as parsing captures, extracting features, and applying learned detectors, yet these do not form a unified action space for an agent. They are scattered across papers; some methods release no implementation, while others provide research prototypes with often incompatible interfaces and dependencies. Even when tools are available, composition remains difficult: isolated calls pass inputs and outputs as text or JSON through the LLM, while in-memory state, such as flow data parsed from a capture, cannot persist across calls. Loops, branches, and other stateful logic therefore require repeated model round trips. Generating free-form glue code is also unreliable: snippets may hallucinate APIs, reference out-of-scope variables, or import unavailable libraries.
As Figure~\ref{fig:challenge_measurement}(b) shows, at most 78\% of
%first-attempt 
code snippets for a simple feature-extraction-and-visualization task execute correctly across the five LLMs, and such errors can amplify through subsequent analysis steps.

\paragraph{C3. Fragile long-horizon analysis} Long-horizon analysis chains multiple tasks over extended traffic periods. For example, investigating suspected data theft may require classifying a week of encrypted flows by application, detecting tunnels among unattributed flows, and correlating implicated hosts' transfers across days to confirm exfiltration. Such analysis stresses both memory and execution. Accumulated evidence can exceed the context window, requiring compression without losing important information. Errors can accumulate and derail subsequent analysis. Traffic also evolves, and distribution shift can make stored evidence and conclusions stale. For example, packet-length patterns learned earlier may no longer identify an application after an update changes its distribution.
As Figure~\ref{fig:challenge_measurement}(c) shows, all five LLMs perform poorly on web attack detection, and their performance degrades further (below 50\% F1) after introducing unseen MAWI~\cite{mawi} background traffic.

%% file: sections/system_design.tex
\section{System Design}

% "fragmented action space" 的整体意象是:动作空间不是一个统一的、可自由组合的整体,而是一堆碎片——能力碎在各个系统里,执行碎在一次次孤立调用里。 这直接反衬 \tool 的两个设计:automated tool construction 把第一层碎片熔成统一工具集(4.3),unified code-execution space 把第二层碎片熔成一个共享状态的执行空间(4.4)。

\subsection{Agent Loop Overview}

\begin{figure}[t]
\centering
\includegraphics[width=0.47\textwidth]{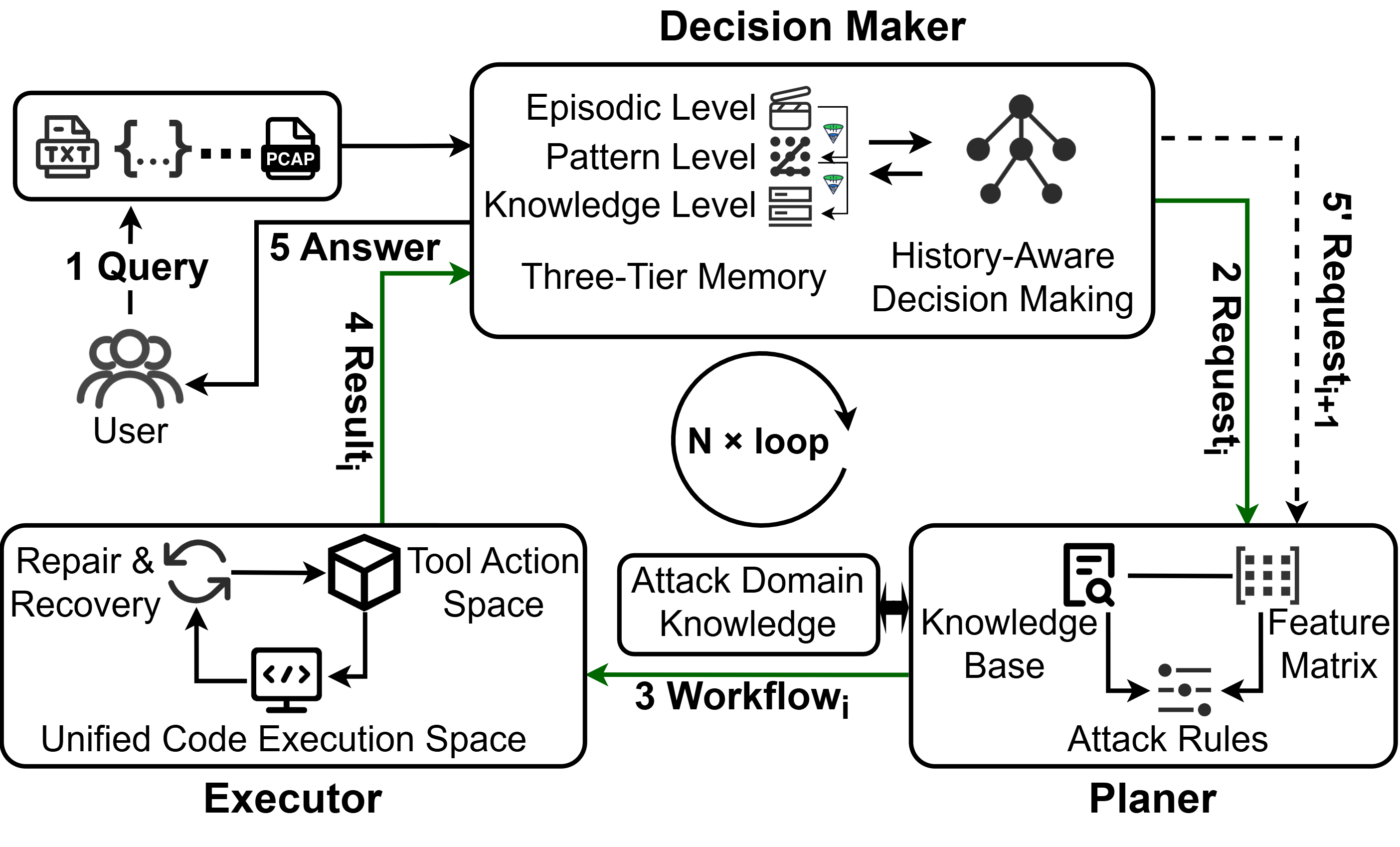}
\caption{The \tool architecture. A decision maker, planner, and executor form an agent loop, supported by knowledge-augmented workflow planning, a unified code execution space, and three-tier memory for history-aware decision making.}
\label{fig:architecture}
\end{figure}

Figure~\ref{fig:architecture} shows the \tool architecture. \tool takes a natural-language query and traffic capture and returns a conclusion about the traffic. Queries range from simple tasks, such as identifying the application behind a flow, to complex investigations that chain multiple analyses, such as determining whether a host is exfiltrating data through a covert tunnel. \tool answers each query through an iterative \emph{agent loop} with three LLM-based agents: the decision maker, planner, and executor. These agents iteratively plan, execute, and refine the analysis until the query is resolved.

\paragraph{Decision maker} The decision maker consolidates findings across iterations and decides what to do next. At each iteration, it formulates a \emph{planning request} from the user query and findings so far, specifying an analysis objective and the traffic to examine, and sends it to the planner. For example, one request may classify encrypted flows by application, while a later request examines unmatched flows for DNS or TLS tunneling. After a request is planned and executed, the decision maker summarizes the results and either issues the next request, revises the current one, or returns a final conclusion. To support long-horizon decision making (C3), \tool maintains a \emph{three-tier memory} that abstracts evidence into recurring patterns and consolidated long-term knowledge, with relevant entries retrieved to augment decisions.

\paragraph{Planner} The planner converts each planning request into an execution workflow based on the traffic, tool descriptions, and prior request results. A \emph{workflow} is an ordered sequence of \emph{execution steps}, each invoking a verified tool or specifying an operation for the executor to implement with generated code, such as filtering flows or aggregating results. The planner represents step dependencies in a dataflow graph. To bridge the semantic gap (C1), \tool maps high-level attack knowledge rules extracted from the literature to low-level traffic feature conditions. Given a planning request and observed traffic, the planner retrieves relevant knowledge to augment its context and guide workflow generation and tool selection. If a step cannot be completed, the planner regenerates that step and its dependents.

\paragraph{Executor} The executor runs the planner's workflow and returns results to the decision maker. To address C2, \tool introduces two designs to form a unified action space. First, it provides a library of \emph{150+ verified atomic tools} extracted from 50+ published traffic analysis systems and organized into a five-layer hierarchy that constrains valid tool chains. Second, it provides a persistent \emph{code-based execution space}, where actions are composed as executable Python code rather than isolated JSON/text tool calls. Intermediate data stays in memory and passes through variables. Execution runs in a sandbox, where interpreter feedback and the dataflow graph support failure localization, code regeneration, and selective re-execution.

%%%%%%%%%%%%%%%%%%%%%%%%

\subsection{Workflow Planning with Attack Knowledge}
\label{sec:planner}

\begin{figure*}[t]
\centering
\includegraphics[width=.95\textwidth]{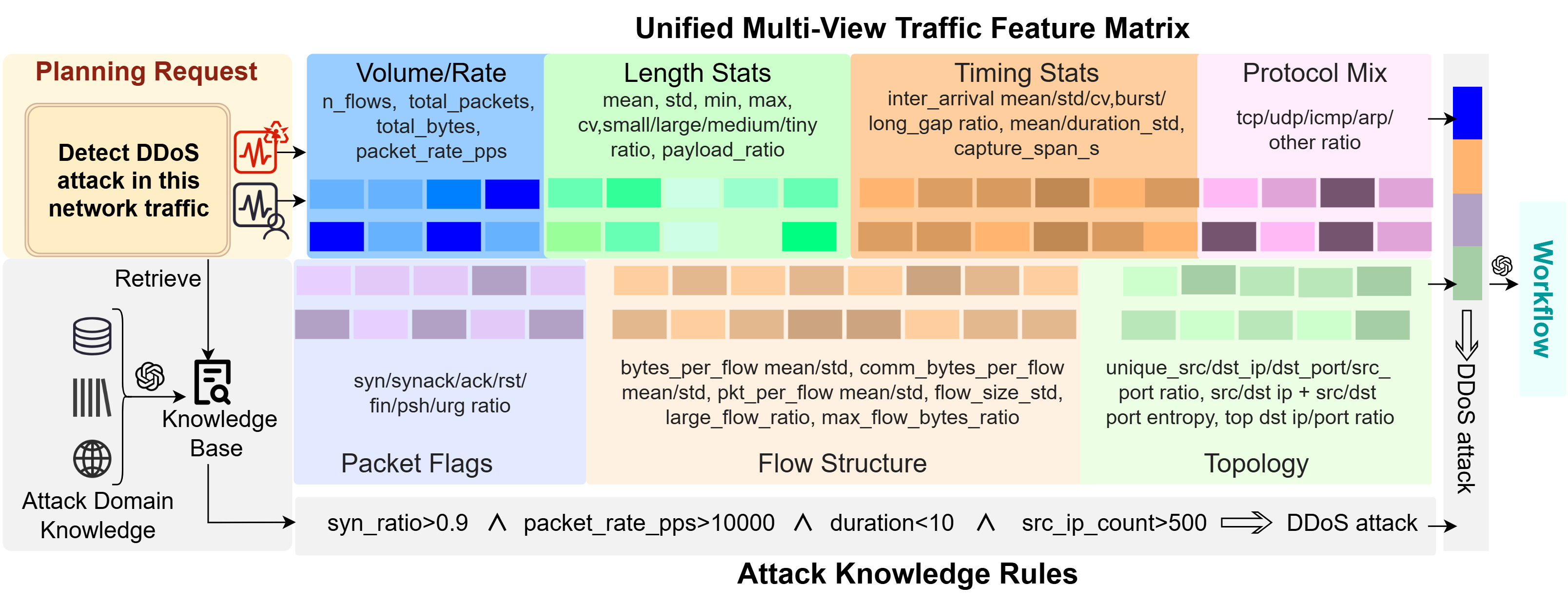}
\caption{Knowledge-augmented workflow planning. \tool bridges the semantic gap using a multi-view traffic representation and an attack knowledge base, with retrieved rules guiding workflow construction and tool selection.}
\label{fig:traffic_matrix}
\vspace{-2ex}
\end{figure*}

\tool bridges the semantic gap (addressing C1) by grounding workflow planning in three components: a unified traffic representation, an attack knowledge base that maps attack patterns to traffic feature conditions, and a retrieval mechanism that augments planning with relevant knowledge (Figure~\ref{fig:traffic_matrix}).

The \textbf{planner} takes a planning request $q$, traffic batch $\mathcal{B}_q$, tool library $\mathcal{T}$, and $H$, the decision maker's conclusions from previous iterations. It outputs a workflow $W=(s_1,\ldots,s_n)$ of execution steps and a dataflow graph $\mathcal{G}$ capturing their dependencies. Each step $s_j$ either invokes a tool $t\in\mathcal{T}$ or specifies an operation for the executor to implement with generated code:
\begin{equation}
(W, \mathcal{G}) = \mathrm{Plan}\big(q,\mathcal{B}_q,\mathcal{T},H,\mathrm{ctx}(q,\mathcal{B}_q)\big)
\label{eq:plan}
\end{equation}
where $\mathrm{ctx}(q,\mathcal{B}_q)$ is knowledge retrieved for the request and observed traffic. The same function supports replanning by incorporating failure feedback and regenerating failed steps and their dependents.

\paragraph{Multi-view traffic representation} Raw captures mix protocols with heterogeneous packet formats that are difficult for an LLM to reason over directly. \tool converts the flows in a capture into a \emph{unified multi-view feature matrix}. 
A view is a feature extractor that maps a flow to one feature value, such as its mean packet length of SYN count. Given $m$ views, a flow $x\in\mathcal{B}_q$ is represented by the row vector:
\begin{equation}
F(x)=\big(f_1(x),\ldots,f_m(x)\big), \quad f_i(x)\in\mathbb{R}
\label{eq:matrix}
\end{equation}
Stacking the row vectors of all flows in $\mathcal{B}_q$ gives the feature matrix $F(\mathcal{B}_q)\in\mathbb{R}^{|\mathcal{B}_q|\times m}$, in which each row corresponds to a flow and each column to a view.
Feature tools in $\mathcal{T}$ compute the views, covering packet-length statistics, inter-arrival times, flow statistics, protocol fields, TCP flags, and packet-length distributions, totaling over 50 views. Payload content is excluded in the current design because much traffic is encrypted and payload features can drift across application versions.

The feature matrix for $\mathcal{B}_q$ is provided to the planner together with a specification that defines each feature's format, meaning, and expected range. For example, the SYN-count specification states that the feature records the number of packets with the SYN flag set and that a normal TCP flow contains one SYN from the client and one SYN-ACK from the server. The planner can therefore interpret an unusually high SYN count as repeated connection attempts, potentially indicating scanning or connection failure. This unified representation gives the planner a consistent semantic space across protocols and tasks.

\paragraph{Attack knowledge base} \tool maintains an \emph{attack knowledge base} $\mathcal{K}$ that links high-level attack knowledge to traffic features. Each rule $(c,r)\in\mathcal{K}$ associates an attack category $c$, such as DDoS, DNS tunneling, or cryptomining, with a first-order logic formula:
\begin{equation}
\forall x.\ \mathit{cond}_1(x)\wedge\cdots\wedge\mathit{cond}_k(x)\Rightarrow c(x)
\label{eq:rule}
\end{equation}
together with a natural-language explanation. Each $\mathit{cond}_i(x)$ is a predicate over a traffic feature, such as a threshold or range check.
For example, a cryptomining rule may be:
\begin{equation*}
\begin{gathered}
\forall x.\ \mathrm{Dur}(x)>600 \wedge \mathrm{MeanPktLen}(x)<200 
\wedge\ \mathrm{MeanIAT}(x)>5 \\ \Rightarrow \mathrm{CryptoMining}(x)
\end{gathered}
\end{equation*}
This rule captures three mining characteristics: long-lived connections, small packets, and large inter-arrival times, reflecting persistent miner-pool connections that exchange short messages with little bulk data.

In the current design, rules are extracted by an LLM from the same 50+ published works used to construct the tool library. Given a paper, the LLM identifies the target attack and the traffic-feature conditions used by the method, then expresses them in the logical form. Each rule is manually verified against the source paper before being added to $\mathcal{K}$. The same process can also extend the knowledge base with other sources, such as public threat reports.

\paragraph{Retrieval-augmented planning} Given the request $q$ and the observed traffic $\mathcal{B}_q$, the planner retrieves attack rules to augment workflow generation. The goal is to retrieve rules for the attack or behavior targeted by the request. Let $\phi$ denote a text embedding model and $\mathrm{sim}$ cosine similarity. Each rule $(c,r)$ is embedded once from its attack category and formula, and the query embeds $q$ together with the batch representation $F(\mathcal{B}_q)$. The planner retrieves the $k_r$ most similar rules:

\begin{equation}
\mathrm{ctx}(q,\mathcal{B}_q)=
\operatorname*{arg\,top\text{-}k_r}_{(c,r)\in\mathcal{K}}\;\;
\mathrm{sim}\big(\phi(c,r),\phi(q,F(\mathcal{B}_q))\big)
\label{eq:retrieval}
\end{equation}
These rules and their explanations form the $\mathrm{ctx}(q,\mathcal{B}_q)$ term in Equation~\ref{eq:plan}. Their conditions identify which traffic features must be computed, guiding tool selection and workflow construction. For example, when analyzing suspicious DNS traffic, the planner may retrieve a DNS tunneling rule whose conditions specify features such as query length and frequency, and generate workflow steps to compute them.

%%%%%%%%%%%%%%%%%%%%%%%%

\subsection{Tool Action Space Construction}
\label{sec:tools}

Traffic analysis requires diverse capabilities, yet no unified action space exists for the traffic domain. These capabilities are scattered across published papers, with many lacking released implementations and others providing research prototypes with incompatible interfaces. Without such tools, an agent cannot effectively act on its environment. Our \emph{insight} is that diverse traffic analysis tasks share reusable operations that can be extracted once and dynamically composed into workflows for different tasks. We therefore construct a unified action space as \tool's tool library $\mathcal{T}$ (addressing C2), where each tool is a Python function with a typed signature and natural-language specification.

\paragraph{Automatic tool discovery from literature} We use an LLM-based pipeline to mine reusable operations from 50+ traffic analysis systems published at top-tier security and networking venues, covering the task families in Section~\ref{subsec:tasks}. For each paper, the pipeline identifies its target task, decomposes the method into operations using the five-layer taxonomy in Section~\ref{sec:executor}, and assigns each operation to a layer. When source code is available, we adapt the corresponding implementation. Otherwise, we implement the tool from the paper description. Each resulting tool exposes a uniform typed interface and natural-language specification and is traceable to its source paper and operation. To ensure functional correctness, each implementation is manually verified against the paper, tested on predefined cases, and, when applicable, validated by reproducing the reported results on the same datasets before being finalized into the library. As Table~\ref{tab:extraction_ledger_9tasks} in Appendix~\ref{app:results} shows, the reproduced performance is comparable to the original results.

\paragraph{Coverage-preserving tool selection} Tool discovery may yield overlapping candidates because source systems often reimplement the same operations with minor variations. Retaining them can confuse the planner and waste computation. Let $\mathcal{C}$ denote the candidate set, $\mathcal{U}=\{u_1,\ldots,u_n\}$ the discovered operations, and $\mathrm{cov}(t)\subseteq\mathcal{U}$ the operations implemented by candidate $t$. We seek the smallest subset that preserves full coverage:
\begin{equation}
\min_{\mathcal{T}\subseteq\mathcal{C}} |\mathcal{T}| \quad
\text{s.t.} \quad
\bigcup_{t\in\mathcal{T}}\mathrm{cov}(t)=\mathcal{U}
\label{eq:setcover}
\end{equation}
which is the NP-hard set cover problem. We approximate it with a greedy pruning algorithm %(Algorithm~\ref{alg:toolselect} in \confirm{Appendix~\ref{x}})
that repeatedly selects the candidate covering the most uncovered operations.
%, breaking ties in favor of the \confirm{simplest interface, measured by the fewest arguments and return values}. 
After full coverage is reached, a pruning pass removes any selected tool whose operations remain covered by others.
We then manually review the selected set to remove redundancy that the algorithm cannot detect, such as semantically identical operations recorded under different names, and to exclude trivial operations that can be implemented with a few lines of generated code and do not warrant a dedicated tool. Together, these steps remove over 85\% of the original candidates, yielding 157 tools in the final library $\mathcal{T}$. Table~\ref{tab:tools} in Appendix~\ref{app:results} shows the list of tools, covering critical capabilities across diverse traffic analysis tasks.

%%%%%%%%%%%%%%%%%%%%%%%%

\begin{table}[t]
\centering
\caption{Five functional layers of the \tool tool library.}
\label{tab:layers}
\small
\setlength{\tabcolsep}{4pt}
\begin{tabular}{@{}p{1.4cm}p{\dimexpr\columnwidth-1.4cm-2\tabcolsep\relax}@{}}
\toprule
\textbf{Layer} & \textbf{Contents} \\
\midrule
L1 & \textbf{Data Reading:} Reads and preprocesses heterogeneous inputs, including PCAP packet captures, NetFlow records, and CSV or TXT feature tables.\\
L2 & \textbf{Feature Extraction:} Extracts atomic per-packet and per-flow features, such as packet lengths, inter-arrival times, header fields, and payload entropy.\\
L3 & \textbf{Feature Construction:} Builds composite features from atomic ones, such as statistics over a time window, multi-flow and per-host distributions, and learned representations from pre-trained encoders. \\
L4 & \textbf{Attack Characterization:} Derives attack-specific features based on augmented knowledge, such as field-value ranges, beaconing periodicity, and anomalous interaction patterns.\\
L5 & \textbf{Analysis Models:} Applies learned models, including supervised classifiers, anomaly detectors, and clustering, to produce traffic analysis results. \\
\bottomrule
\end{tabular}
\end{table}

\subsection{Unified Code Execution Space}
\label{sec:executor}

Prior agent designs commonly invoke tools through isolated JSON or text calls, passing intermediate results back through the model with no persistent state. As a result, data transfer, loops, branches, and intermediate computation require repeated model round trips.

\paragraph{Code as the composition medium} \tool instead uses generated Python code as the composition layer between tools. The executor maintains a persistent \emph{code space} $N$, a Python namespace containing all tools in $\mathcal{T}$ as callable functions and variables produced by earlier steps. Each workflow step is realized as a Python snippet that executes in $N$, invokes tools, and stores outputs as variables that later steps can reference directly. This design offers several benefits. A snippet can compose multiple tools with loops, branches, and intermediate computations. Data passes directly in memory rather than through the model context, reducing token overhead and avoiding serialization and type-conversion errors. Tools can also be combined seamlessly with standard Python libraries without requiring separate wrappers.

The \textbf{executor} receives the workflow $W=(s_1,\ldots,s_n)$ and dataflow graph $\mathcal{G}$ from the planner. Each step $s_j=(\mathit{act}_j,\mathit{args}_j,\mathit{deps}_j)$ specifies an action, either a tool in $\mathcal{T}$ or an operation implemented with generated code, its arguments $\mathit{args}_j$, and dependencies $\mathit{deps}_j$ on prior outputs. The executor generates a snippet $c_j$ and executes it in the current code space $N_j$:
\begin{equation}
(r_j,N_{j+1})=\mathrm{Exec}(s_j,N_j)
\label{eq:exec}
\end{equation}
where $N_1$ contains the tools in $\mathcal{T}$ and planning-request inputs. The result $r_j$ is bound in $N_{j+1}$ under the step's name for later step's reference. After the final step, the executor returns $(r_1,\ldots,r_n)$ to the decision maker.

\paragraph{Layered tool composition} The flexibility of the code space can lead the agent to chain tools in invalid orders. A snippet may feed a classifier features that were never extracted, or assume that feature extraction subsumes packet parsing and omit the parsing step. Without safeguards, such errors surface only at run time. \tool prevents this by organizing tools into five layers, $L_1$ through $L_5$ (Table~\ref{tab:layers}), following the traffic analysis pipeline from raw-data reading to classification, and enforcing a partial order over tool composition.

We define $L_i \prec L_j$ to mean that outputs from tools in $L_i$ may feed tools in $L_j$. The order follows $L_1 \prec L_2 \prec \cdots \prec L_5$, and reverse composition is disallowed. For example, a feature extractor may consume parsed packets, but a parser cannot consume extracted features. Let $L(t)$ denote the layer of tool $t$, and $t \triangleright t'$ indicate that an output of $t$ is passed to $t'$. The set of valid tool chains is:
\begin{equation}
\mathrm{Chains}(\mathcal{T}) =
\big\{
t_1 \triangleright \cdots \triangleright t_p
\mid
t_i \in \mathcal{T},\
L(t_1) \preceq \cdots \preceq L(t_p)
\big\}
\label{eq:flows}
\end{equation}
where layers may be skipped when unnecessary, so chains such as $L_1 \rightarrow L_4$, $L_1 \rightarrow L_5$, and $L_1 \rightarrow L_3 \rightarrow L_5$ are valid. For example, signature matching may parse a capture at $L_1$ and directly match packet contents at $L_4$, while a classifier may require parsing at $L_1$, feature extraction at $L_2$, and classification at $L_5$. The planner checks this ordering when constructing the workflow, catching invalid or missing steps before execution. Within each layer, tools expose uniform interfaces, allowing them to be substituted or run in parallel on the same input, for example by applying several feature extractors to the same parsed flows.

%%%%%%%%%%%%%%%%%%%%%%%%

\subsection{Reliable Execution and Failure Recovery}
\label{sec:recovery}

\begin{figure}[t]
\centering
\includegraphics[width=\linewidth]{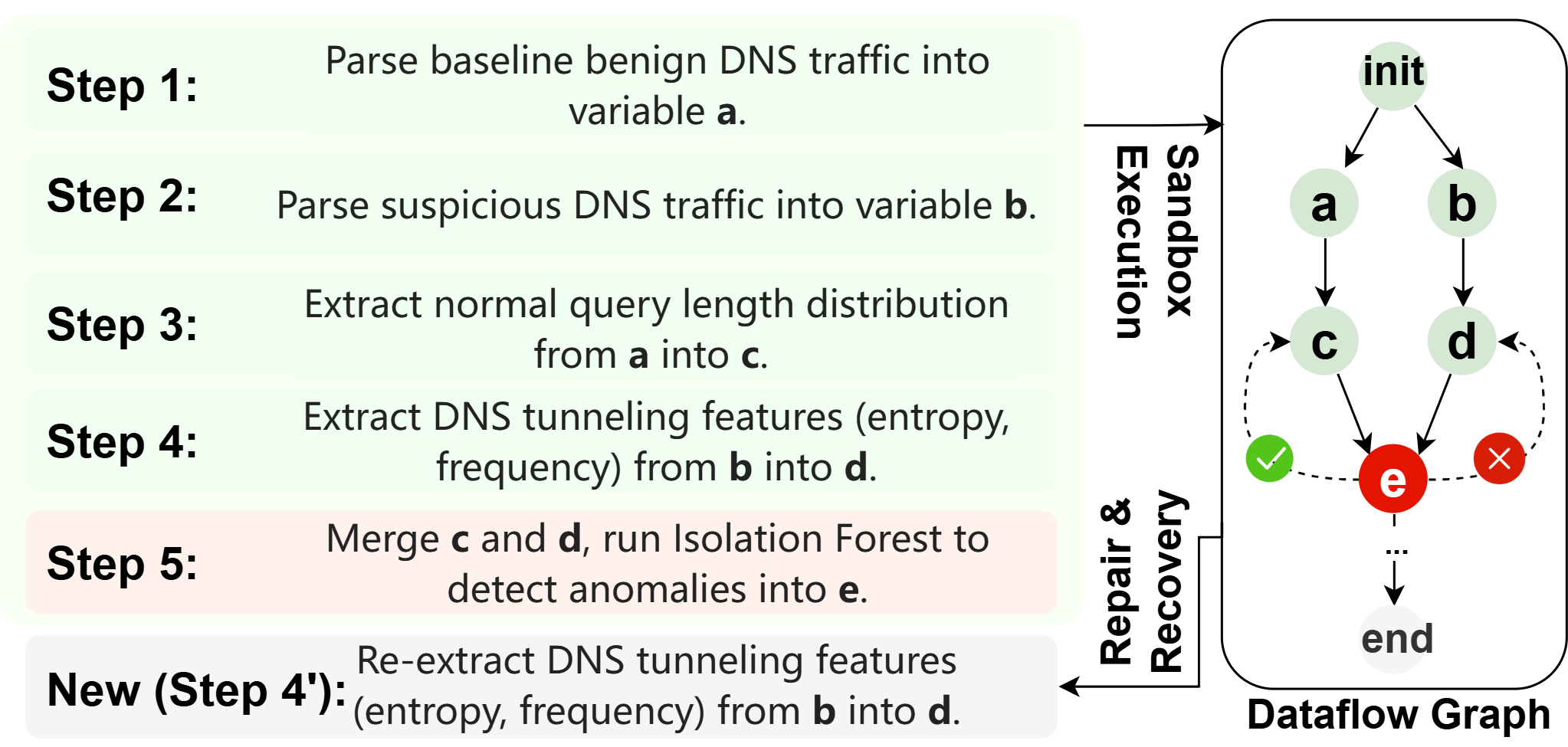}
\caption{\toolnospace’s failure recovery mechanism. Sandbox feedback diagnoses failures, and the dataflow graph enables selective re-execution of affected steps for reliable runtime repair.}
\label{fig:fail_rec}
\vspace{-2ex}
\end{figure}

% Each step runs in its own sandbox, generated code passes static checks before execution and runtime guardrails during execution, and a failed step is repaired and re-executed together with only the steps that depend on it.

Run-time code generation can introduce reliability problems, such as nonexistent APIs, invalid variables, incompatible data formats, or partial state left by failed steps. \tool therefore adds several mechanisms to the code space for reliable execution and recovery.

\paragraph{Per-step sandboxed execution} To isolate errors and keep the code space clean, each step runs in a fresh lightweight sandbox that is destroyed after completion. Only explicitly recorded results are committed to the persistent code space $N$; temporary variables, intermediate files, loaded models, and other working state are discarded. This prevents failed steps from leaving partial state and keeps $N$ compact. Within the sandbox, file access follows a three-zone model. The input zone is read-only and contains the original traffic data and other verified inputs. The private compute zone processes data from the input zone and is destroyed after the step. The output zone is writable only from the compute zone and stores artifacts committed to $N$. This separation protects the source data and persistent code space while ensuring that only necessary outputs persist.

\paragraph{Pre-execution checking} Before execution, the executor validates each snippet against the tool signatures and current code space. Tool calls are checked for required arguments and type consistency. Each referenced variable is verified to exist in $N$ and contain valid data. \tool also maintains an import whitelist consisting of the standard library and 20 approved analysis libraries, and rejects unsafe operations such as \texttt{eval}, dynamic imports, and arbitrary file access. Snippets that fail these checks are regenerated before execution.

\paragraph{Runtime repair and recovery} A snippet that passes pre-execution checks may still fail at run time, for example when a tool encounters unexpected data and raises an exception. \tool introduces a mechanism for runtime repair and dynamic recovery. The sandbox captures the Python interpreter feedback and exceptions, together with the log messages of the step, and sends this information to the executor to locate the failure and diagnose its cause. The executor then regenerates the snippet using the diagnosed cause, step objective, tool signatures, and currently available variables, and re-runs it in a fresh sandbox with the same inputs. If the failure originates upstream, the executor uses the dataflow graph $\mathcal{G}$ to identify the faulty step and re-executes only that step and its dependents in dependency order. For example, a classifier may fail because an earlier step produced feature vectors with fewer dimensions than expected. Results from unaffected steps remain unchanged in $N$. If recovery still fails, the executor reports the failure and state of $N$ to the planner, which regenerates the failed step and its dependents using alternative tools. Figure~\ref{fig:fail_rec} illustrates this recovery process.

%%%%%%%%%%%%%%%%%%%%%%%%

\subsection{Three-Tier Memory}
\label{sec:memory}

Over a long horizon, the decision maker receives a sequence of user queries over evolving traffic, each asking for one traffic analysis task or several when the query is vague. Earlier conclusions may inform later queries, but network drift can make them stale or conflicting. The decision maker therefore tracks and consolidates results and conclusions across the horizon to make informed decisions.

\paragraph{History-aware decision making} The \textbf{decision maker} answers each query $Q$ over agent-loop iterations $t=1,2,\ldots$. At iteration $t$, it receives the planning request $q_t$, executor results $R_t=(r_1,\ldots,r_n)$, prior conclusions $H_t=(h_1,\ldots,h_{t-1})$, and memory entries relevant to $Q$. It produces the current conclusion $h_t$, confidence $p_t\in[0,1]$, and next action $a_t$, either another planning request or the final answer to $Q$:
\begin{equation}
(h_t,p_t,a_t)=\mathrm{Dec}\big(Q,q_t,R_t,H_t,\mathrm{Ret}(Q,\mathcal{M})\big)
\label{eq:decide}
\end{equation}
where $\mathrm{Dec}$ is the decision maker and $\mathrm{Ret}$ the memory retrieval function. The conclusion $h_t$ is appended to $H_t$ and written to memory with $q_t$, its timestamp, and $p_t$.

The decision maker retrieves memory entries that are both relevant and confident:
\begin{equation}
\mathrm{Ret}(Q,\mathcal{M})=
\operatorname*{arg\,top\text{-}k_m}_{e\in\mathcal{M}}\;\;
\mathrm{sim}(Q,e)\cdot w_e
\label{eq:memret}
\end{equation}
where $e$ is a memory entry, $\mathrm{sim}(Q,e)$ denotes its semantic similarity to $Q$, $w_e$ is the entry weight defined below, and $k_m$ is the number of entries retrieved.

\paragraph{Three-level memory and consolidation} The memory $\mathcal{M}=(\mathcal{E},\Sigma,\Pi)$ has three levels and progressively compresses conclusions. The \emph{episodic} level $\mathcal{E}$ stores per-iteration conclusions with timestamps. The \emph{pattern} level $\Sigma$ stores key conclusions merged from similar episodes. The \emph{knowledge} level $\Pi$ stores long-term conclusions, such as routine host behavior or recurring attack patterns. When the decision maker produces the conclusion $h_t$ with confidence $p_t$, the conclusion is added to the episodic level as an entry $e$ with its timestamp, and the entry receives a weight that balances its novelty and confidence:
\begin{equation}
w_e=\sigma\big(\alpha\cdot\mathrm{Novelty}(e\mid\mathcal{E})+(1-\alpha)\cdot p_t\big)
\label{eq:weight}
\end{equation}
where $\sigma$ is the sigmoid function, $\alpha \in [0, 1]$ balances the two terms, and $\mathrm{Novelty}(e\mid\mathcal{E})$ is the rarity of the conclusion among the existing episodes, estimated from a histogram density, so that a conclusion unlike previous ones receives a higher weight.

Consolidation proceeds upward.  As episodes accumulate, similar entries in $\mathcal{E}$ are merged into patterns in $\Sigma$. For example, six consecutive ten-minute episodes reporting SYN flood packets from a host to the same server are merged into one pattern spanning 0:00 to 1:00. When a pattern recurs beyond a threshold, it is distilled into a knowledge entry in $\Pi$. Distillation can also combine conclusions across queries. For example, an intrusion detection query may find that a host contacts an external domain every minute with a few small packets, and a later application classification query finds that these flows match no known application. Together, they yield the knowledge that the host runs an unknown program beaconing to that domain. Episodic entries use $w_e$ from Equation~\ref{eq:weight}. Pattern and knowledge entries inherit the mean weight of their source entries.

Memory entries may become stale or conflicting due to network drift. During consolidation, an entry $e$ is retained only if:
\begin{equation}
\mathrm{Age}(e)<\tau_{\mathrm{age}}
\;\wedge\;
\mathrm{Freq}(e)>\tau_{\mathrm{freq}}
\;\wedge\;
\mathrm{Conflict}(e)<\tau_{\mathrm{conflict}}
\label{eq:keep}
\end{equation}
where $\mathrm{Age}(e)$ is the time since $e$ was created or last updated, $\mathrm{Freq}(e)$ is the number of conclusions matching $e$, and $\mathrm{Conflict}(e)$ is the number of knowledge entries it contradicts. $\tau_{\mathrm{age}}$, $\tau_{\mathrm{freq}}$, and $\tau_{\mathrm{conflict}}$ are thresholds. Recent and recurring entries are retained, while stale or conflicting ones are removed. If further compression is needed, the lowest-weight entries are removed next. This design allows memory to adapt to distribution shift by replacing outdated conclusions with newer, recurring ones.

% \begin{figure}[t]
% \centering
% \includegraphics[width=0.5\textwidth]{pic/three-layer.drawio.png}
% \caption{Three-Layer Memory}
% \label{fig:memory}
% \end{figure}

%% file: sections/evalution.tex
%-------------------------------------------------------------------------------
\section{Evaluation}
\label{sec:eva}
%-------------------------------------------------------------------------------

We systematically evaluate \tool's performance, generalization, and robustness on 9 widely used traffic analysis benchmarks against 23 single-task and 5 multi-task baselines. Our evaluation addresses the following research questions:

\begin{itemize}[leftmargin=*, itemsep=1pt, topsep=1pt]

    \item \textbf{RQ1. Performance.} How does \tool compare with single-task and multi-task baselines across diverse traffic analysis tasks? 
    \item \textbf{RQ2. Generalization.} How well does \tool generalize to unseen traffic distribution?
    \item \textbf{RQ3. Robustness.} How does \tool perform under the distribution shift caused by real-world background traffic?
    \item \textbf{RQ4. Ablation.} How much does each component, including LLM choice, contribute to \tool's performance?
    
\end{itemize}

%%%%%%%%%%%%%%%%%%%%%

\subsection{Experimental Setup}
\label{sec:exe_set}

\begin{table}[t]
\centering
\caption{Traffic analysis tasks and datasets used for evaluation.}
\label{tab:datasets}

\scriptsize
\setlength{\tabcolsep}{3pt}
\renewcommand{\arraystretch}{1.12}

\begin{tabularx}{\columnwidth}{
    @{}
    >{\raggedright\arraybackslash}p{2.25cm}
    >{\raggedright\arraybackslash}p{1.75cm}
    >{\centering\arraybackslash}p{0.75cm}
    >{\raggedright\arraybackslash}X
    @{}
}
\toprule
\textbf{Task}
& \textbf{Dataset}
& \textbf{Records}
& \textbf{Description}
\\
\midrule

Web attack detection
& CIC-IDS2017~\cite{cicids}
& 2.8M
& 14 attacks and benign traffic.
\\

\rowcolor{gray!7}
Fine-grained attack classification
& CIC-IDS2018~\cite{cicids18}
& 16.2M
& 50 attackers with strong class imbalance.
\\

Network intrusion detection
& UNSW-NB15~\cite{unswnb15}
& 2.5M
& Hybrid traffic with 9 attack families.
\\

\rowcolor{gray!7}
IoT intrusion detection
& CICIoT2023~\cite{ciciot}
& 46.8M
& 33 attacks across 7 categories from 105 devices.
\\

DDoS attack detection
& CIC-DDoS2019~\cite{cicddos2019}
& 50M+
& 12 attacks, including TCP, UDP, and HTTP floods.
\\

\rowcolor{gray!7}
VPN traffic classification
& ISCX-VPN2016~\cite{iscxvpn}
& 9.8K
& 7 applications with VPN and non-VPN traffic.
\\

Website fingerprinting
& DF2018~\cite{df2018}
& 140K
& 95 Tor sites with 41K open-world traces.
\\

\rowcolor{gray!7}
BGP anomaly detection
& CAIDA AS~\cite{caidaas}
& 70M+
& AS-level topology data from 2025.
\\

DGA malicious domain detection
& DGArchive~\cite{dgarchive}
& 100M+
& 93+ families with character- and dictionary-based DGAs.
\\

\rowcolor{gray!7}
Real-world background traffic
& MAWI/WIDE~\cite{mawi}
& 200M+
& Real 10-Gbps backbone traces.
\\

\bottomrule
\end{tabularx}
\end{table}

\paragraph{Tasks and datasets} We evaluate nine traffic analysis tasks (Table~\ref{tab:datasets}): network intrusion detection (UNSW-NB15~\cite{unswnb15}), web attack detection (CIC-IDS2017~\cite{cicids}), fine-grained attack classification (CIC-IDS2018~\cite{cicids18}), DDoS detection (CIC-DDoS2019~\cite{cicddos2019}), IoT intrusion detection (CICIoT2023~\cite{ciciot}), VPN traffic classification (ISCX-VPN2016~\cite{iscxvpn}), website fingerprinting (DF2018~\cite{df2018}), BGP anomaly detection (CAIDA~\cite{caidaas}), and DGA malicious domain  detection (DGArchive~\cite{dgarchive}). All are widely used public benchmarks with extensive prior work, providing mature baselines for systematic comparison. For robustness evaluation, we additionally use real-world traffic from MAWI/WIDE~\cite{mawi}, an archive of daily packet traces collected on the transit link between Japan's WIDE backbone and its upstream ISP.

%\confirm{archive of backbone traffic collected from Japanese network research and academic institutions.}

\paragraph{Baselines}
We compare \tool against 28 baselines: 23 single-task and 5 multi-task methods. The single-task baselines include 7 classical ML methods (Random Forest, SVM, KNN, Gradient Boosting, Logistic Regression, Decision Tree, and KMeans) and 16 SOTA deep learning methods: HyperVision~\cite{hypervision}, RAPIER~\cite{rapier}, PointCloud~\cite{pointcloud}, Lemon~\cite{lemon}, NetRadar~\cite{netradar}, Rosetta~\cite{rosetta}, TFE-GNN~\cite{tfegnn}, Sugar~\cite{sugar}, Deep Fingerprinting~\cite{df2018}, Swallow~\cite{swallow}, Multi-tab WF~\cite{multitabwf}, IoT Device ID~\cite{iotdeviceid}, WiFinger~\cite{wifinger}, Semantics-Aware~\cite{semanticsaware}, Joint Adversarial Training DGA Classification~\cite{robustdga}, and IDS-Agent~\cite{idsagent}. 
The multi-task baselines are nPrint~\cite{nprint}, ET-BERT~\cite{etbert}, TrafficFormer~\cite{trafficformer}, TrafficLLM~\cite{trafficllm}, and a ReAct agent~\cite{yao2022react} equipped with the \tool tool library.

\paragraph{Configurations and metrics}
Single-task baselines are evaluated on their target datasets, and multi-task baselines across all nine datasets. We use released code and default settings when available. Otherwise, we reimplement and verify the method from the paper. \tool and ReAct use GPT-5.5 as the backbone LLM; Section~\ref{sec:abl} evaluates alternative LLMs from 9B to 284B parameters. Each dataset is split 80\%/20\% for training and testing, with results averaged over 10-fold cross-validation. Table~\ref{tab:hyperparameters} in Appendix~\ref{app:results} lists \tool's configuration parameters. We use F1 score as the primary task metric.

%%%%%%%%%%%%%%%%%%%%%

\subsection{RQ1: Performance} 
\label{sec:eval_rq1}

\begin{table*}[t]
\centering
\caption{Performance comparison across evaluation benchmarks. For each task, we report \tool's end-to-end F1, the best single-task and multi-task baselines, and the corresponding absolute F1 differences ($\Delta_S$ and $\Delta_M$).}
\label{tab:benchmarks}

\small
\setlength{\tabcolsep}{4pt}
\renewcommand{\arraystretch}{1.15}

\begin{adjustbox}{width=.95\textwidth}

\begin{tabularx}{\textwidth}{
    @{}
    >{\raggedright\arraybackslash}p{3.45cm}
    >{\raggedright\arraybackslash}p{2.55cm}
    >{\centering\arraybackslash}p{1.20cm}
    >{\raggedright\arraybackslash}X
    >{\centering\arraybackslash}p{1.15cm}
    >{\raggedright\arraybackslash}X
    >{\centering\arraybackslash}p{1.15cm}
    @{}
}
\toprule

\multirow{2}{*}{\textbf{Task}}
&
\multirow{2}{*}{\textbf{Benchmark}}
&
\multirow{2}{*}{\textbf{\tool}}
&
\multicolumn{2}{c}{\textbf{Best Single-Task}}
&
\multicolumn{2}{c}{\textbf{Best Multi-Task}}
\\

\cmidrule(lr){4-5}
\cmidrule(lr){6-7}

&
&
&
\textbf{Method / F1}
&
$\boldsymbol{\Delta_S}$
&
\textbf{Method / F1}
&
$\boldsymbol{\Delta_M}$
\\

\midrule

\rowcolor{gray!7}
Network intrusion detection
& UNSW-NB15~\cite{unswnb15}
& \textbf{0.9863}
& \mbox{PointCloud~\cite{pointcloud}, 0.9750}
& $+1.13\%$
& \mbox{nPrint~\cite{nprint}, 0.9807}
& $+0.56\%$
\\

Web attack detection
& CIC-IDS2017~\cite{cicids}
& \textbf{0.9995}
& \mbox{HyperVision~\cite{hypervision}, 0.9920}
& $+0.75\%$
& \mbox{TrafficLLM~\cite{trafficllm}, 0.9955}
& $+0.40\%$
\\

\rowcolor{gray!7}
Fine-grained attack classification
& CIC-IDS2018~\cite{cicids18}
& \textbf{0.9999}
& \mbox{Classical ML~\cite{leevy2021}, 0.9999}
&  0
& \mbox{TrafficFormer~\cite{trafficformer}, 0.9985}
& $+0.14\%$
\\

IoT intrusion detection
& CICIoT2023~\cite{ciciot}
& \textbf{0.9998}
& \mbox{IDS-Agent~\cite{idsagent}, 0.9691}
& $+3.07\%$
& \mbox{ReAct-Agent~\cite{yao2022react}, 0.8262}
& $+17.36\%$
\\

\rowcolor{gray!7}
DDoS detection
& CIC-DDoS2019~\cite{cicddos2019}
& \textbf{0.9999}
& \mbox{NetRadar~\cite{netradar}, 0.9989}
& $+0.10\%$
& \mbox{nPrint~\cite{nprint}, 0.9850}
& $+1.49\%$
\\

VPN classification
& ISCX-VPN2016~\cite{iscxvpn}
& 0.9242
& \mbox{Rosetta~\cite{rosetta}, 0.9120}
& $+1.22\%$
& \mbox{ET-BERT~\cite{etbert}, 0.9454}
& $-2.12\%$
\\

\rowcolor{gray!7}
Website fingerprinting
& DF2018~\cite{df2018}
& \textbf{0.9965}
& \mbox{DeepFP~\cite{df2018}, 0.9797}
& $+1.68\%$
& \mbox{TrafficFormer~\cite{trafficformer}, 0.9520}
& $+4.45\%$
\\

BGP anomaly detection
& CAIDA AS~\cite{caidaas}
& 1.00
& \mbox{Semantics-Aware~\cite{semanticsaware}, 1.00}
& 0
& \mbox{ReAct-Agent\cite{yao2022react}, 0.9752}
& $+2.48\%$
\\

\rowcolor{gray!7}
DGA detection
& DGArchive~\cite{dgarchive}
& \textbf{0.9999}
& \mbox{Classical ML, 0.9899}
& $+1.00\%$
& \mbox{nPrint~\cite{nprint}, 0.9750}
& $+2.49\%$
\\

\bottomrule
\end{tabularx}
\end{adjustbox}
\end{table*}

Table~\ref{tab:benchmarks} reports \tool's end-to-end F1 together with the best single-task and multi-task baselines for each task. \tool achieves the best result on 8 of 9 tasks, with F1 above 98.63\% on each. The only exception is VPN traffic classification, where \tool ranks second with 92.42\% F1 behind ET-BERT at 94.54\%. The reason is that VPN traffic classification relies on subtle byte-level encapsulation patterns. \tool's current tool library does not include a byte-level encoder, to avoid the influence of encryption-induced noise in the payload across tasks. Such an encoder can be added as a tool without changing the framework. Across the other 8 tasks, however, \tool exceeds ET-BERT by 35.51 F1 points on average.

Multi-task baselines perform well only on subsets of tasks. nPrint reaches 98.07\% F1 on intrusion detection but only 54.78\% on VPN traffic classification, as its fixed representation and shallow classifiers fail to capture the required latent traffic patterns. 
TrafficFormer and ET-BERT achieve only 35.56\% and 25.88\% F1 on DGA malicious domain detection, as their architectures and pre-training objectives are designed to capture traffic semantics and relations among packet metadata rather than the structure of domain names. TrafficLLM achieves 52.57\% F1 on the same task because the LLM alone cannot identify new malicious domain patterns. The ReAct agent, despite using the same tool library as \tool, remains below 49.97\% F1 on website fingerprinting, VPN classification, and DGA detection. Further analysis shows that it typically invokes no more than three tools to collect basic statistics before drawing a conclusion. Such a general-purpose agent lacks the capability to identify the right tools for a given traffic analysis task and orchestrate the deeper tool chains required for complex analyses.

\begin{takeaway}
    \textbf{Takeaway:} \tool's agentic design provides a more effective multi-task solution, outperforming existing multi-task and single-task approaches across diverse traffic analysis tasks.
\end{takeaway}

\begin{figure}[t]
\centering
\includegraphics[width=\columnwidth]{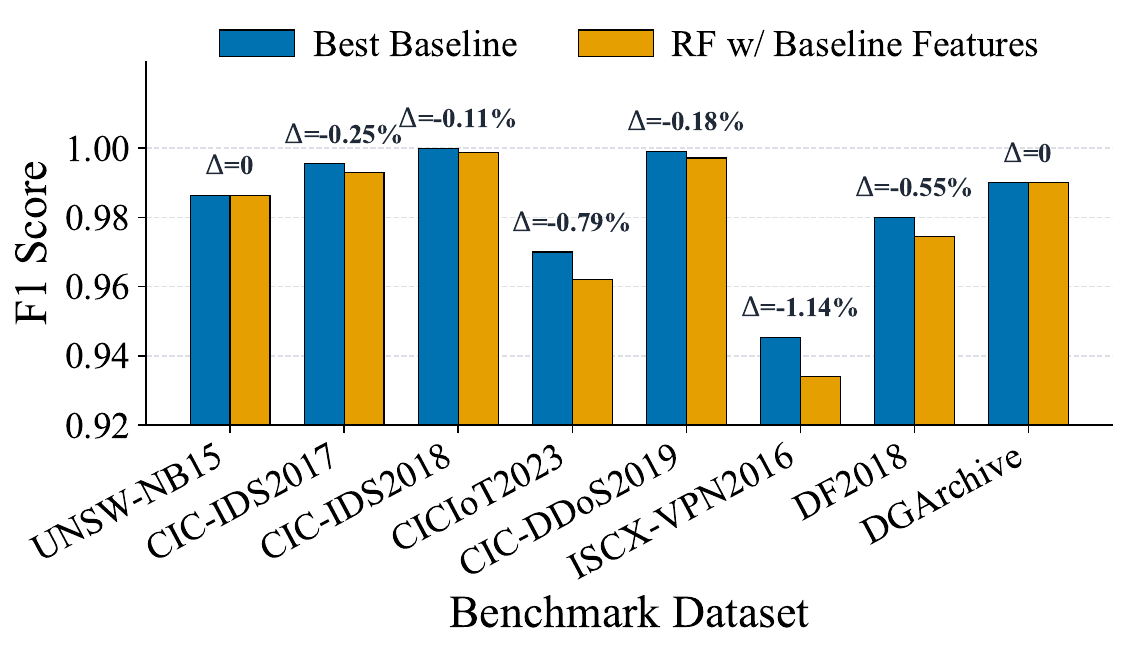}
% \caption{\fh{Comparing the best baseline's F1 with a shared Random Forest on its extracted features. $\Delta$ denotes the F1 difference (Shared RF minus Best Baseline)}}
\caption{Best baseline vs. Random Forest using the same extracted features. $\Delta$ denotes their F1 difference.}
\label{fig:fea-trans}
\vspace{-1ex}
\end{figure}

We also observe that some baselines with similar classification structures can perform very differently on the same task. For example, nPrint and IDS-Agent both use Random Forest, yet differ by 14.59 F1 points on IoT intrusion detection. We hypothesize that feature extraction is a key factor. For each task, we therefore take the features from the best baseline and train a single Random Forest on them. For methods with hand-crafted features, we reproduce them as specified in the original papers. For deep learning methods, we use the output of the last layer before the classification head, such as the encoder output of an autoencoder. As Figure~\ref{fig:fea-trans} shows, this setup stays within 2 F1 points of the best result on every task, indicating that much of the performance comes from the features. Because the best features vary by task, multi-task traffic analysis needs task-specific feature extractors rather than a single fixed representation. This motivates \tool's comprehensive tool library, which includes published feature extractors. The remaining 2-point gap, however, shows that feature extraction with a simple general classifier is not sufficient. \tool's strong performance also relies on its agentic reasoning over intermediate results and decision making.

% \begin{table}[!ht]
% \centering
% \caption{Feature-transfer experiment: for each benchmark, the F1 of the best baseline versus a shared Random Forest trained on features extracted by that baseline's pre-classification network (machine-learning baselines retain their original features).}
% \label{tab:feature}
% \scriptsize
% \setlength{\tabcolsep}{2pt}
% \begin{tabular*}{\columnwidth}{@{\extracolsep{\fill}}lp{2.3cm}ccc}
% \toprule
% \textbf{Benchmark} & \textbf{Feature extractor} & \textbf{Original F1} & \textbf{Shared RF F1} & $\Delta$ \\
% \midrule
% UNSW-NB15 & original features & 0.9863 & 0.9863 & 0 \\
% CIC-IDS2017 & TrafficLLM & 0.9955 & 0.9930 & $-0.25\%$ \\
% CIC-IDS2018 & Leevy & 0.9999 & 0.9988 & $-0.11\%$ \\
% CICIoT2023 & IDS-Agent & 0.9700 & 0.9621 & $-0.81\%$ \\
% CIC-DDoS2019 & HyperVision & 0.9990 & 0.9972 & $-0.18\%$ \\
% ISCX-VPN2016 & ET-BERT & 0.9454 & 0.9340 & $-1.21\%$ \\
% DF2018 & DeepFP & 0.9800 & 0.9745 & $-0.56\%$ \\
% DGArchive & original features & 0.9900 & 0.9900 & 0 \\
% \bottomrule
% \end{tabular*}
% \end{table}

\begin{takeaway}
    \textbf{Takeaway:} Strong multi-task traffic analysis requires both task-specific feature extraction and agentic reasoning. 
\end{takeaway}

%%%%%%%%%%%%%%%%%%%%%%%%%%%%%%%%%
\subsection{RQ2: Generalization}
\label{sec:eval_rq2}

In real-world traffic analysis, deployment traffic may differ substantially from the training distribution even for the same task as applications, services, and user behavior evolve over time. A robust method should generalize under such distribution shifts. We evaluate this for web attack detection using two datasets, UNSW-NB15 and CIC-IDS2017, which share attack categories including DDoS, scanning, reconnaissance, exploits, and infiltration. Both have been widely studied by the community, with baselines achieving 98\% accuracy in-distribution. We train each baseline on UNSW-NB15 and test it on the shared attack categories of CIC-IDS2017.
%, using no CIC-IDS2017 data for training.

\begin{figure}[t]
\centering
\includegraphics[width=\columnwidth]{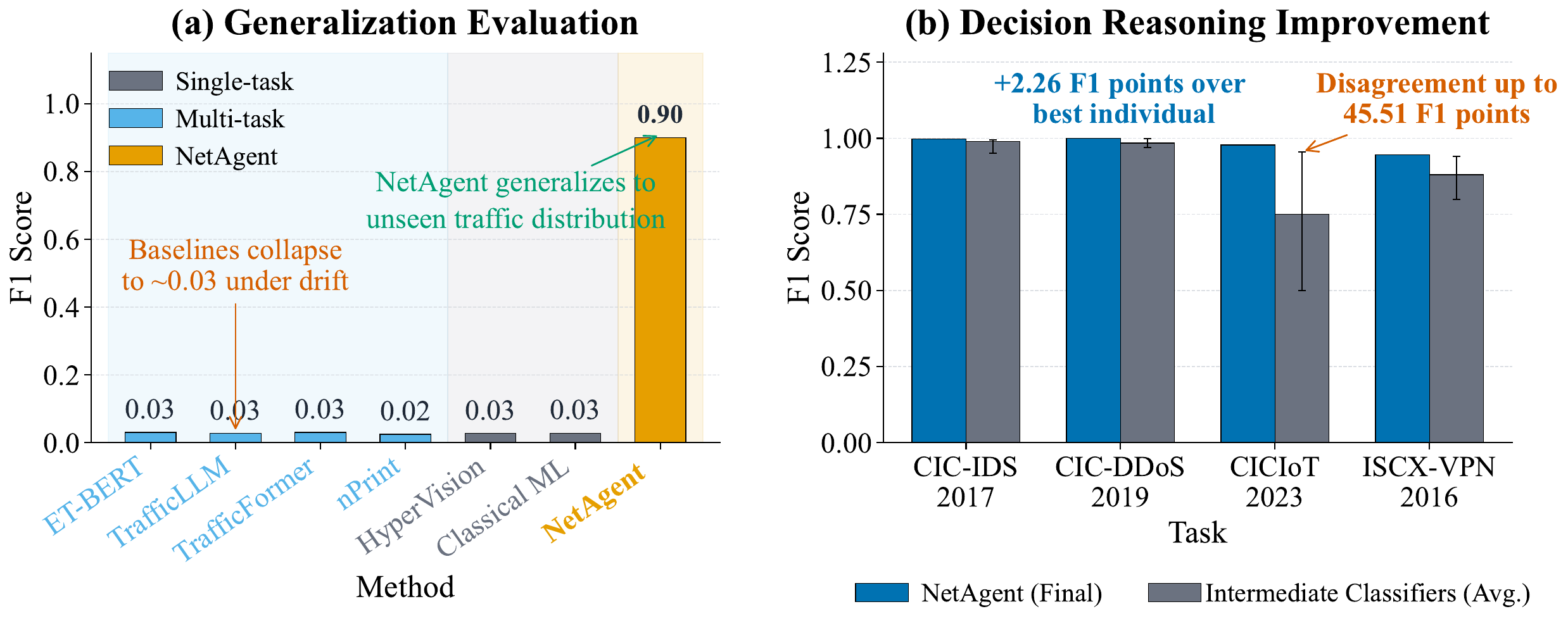}
\caption{(a) Generalization on CIC-IDS2017: baselines drop to about 3\% F1 while \tool achieves 90.04\%. (b) Decision reasoning improves over individual classifier outputs.}
\label{fig:robust-adaptive}
\vspace{-2ex}
\end{figure}

As Figure~\ref{fig:robust-adaptive}(a) shows, all baselines collapse on the unseen distribution. The best single-task and multi-task baselines achieve only 2.74\% and  3.04\% F1, dropping by 95.16 points and 82.56 points from their in-distribution results. In contrast, \tool maintains 90.04\% F1, with a drop of only 9.91 points.

The baselines fail for different reasons. Methods such as nPrint and IDS-Agent rely on shallow classifiers that overfit dataset-specific patterns rather than underlying attack mechanisms, limiting transfer to the same attacks in new datasets. Pre-trained models such as ET-BERT and TrafficFormer underfit benign traffic. Their fine-tuning concentrates on attack patterns in the training data, causing many false positives on benign traffic from the new dataset. TrafficLLM, as a non-agentic fine-tuned LLM, lacks an external knowledge base and remains tied to patterns learned during training, limiting reasoning over unseen traffic. In contrast, \tool retrieves relevant attack knowledge during planning, and its task decomposition over a tool library generalizes effectively to new traffic patterns.

\begin{takeaway}
    \textbf{Takeaway:} \tool generalizes to unseen traffic distributions where all baselines collapse. Planning with retrieved attack knowledge and decomposing the task over a tool library, rather than fitting dataset-specific patterns, enable this generalization.
\end{takeaway}

%%%%%%%%%%%%%%%%%%%%%%%%%

\subsection{RQ3: Robustness}
\label{sec:eval_rq3}

Benchmark traffic is often collected in controlled environments, whereas deployed analyzers must process massive background traffic from live networks. A robust method should therefore distinguish attacks from diverse background noise. We evaluate this on CIC-IDS2017 web attack detection by adding real-world MAWI/WIDE backbone traffic at 5$\times$ the test-set volume. MAWI/WIDE~\cite{mawi} archives daily traffic from the transit link between Japan's WIDE backbone and its upstream ISP, covering diverse protocols and network behaviors. The injected traffic contains over 283 million packets. CIC-IDS2017 labels remain unchanged, and all added flows are treated as benign.

\begin{figure}[t]
\centering
\includegraphics[width=\columnwidth]{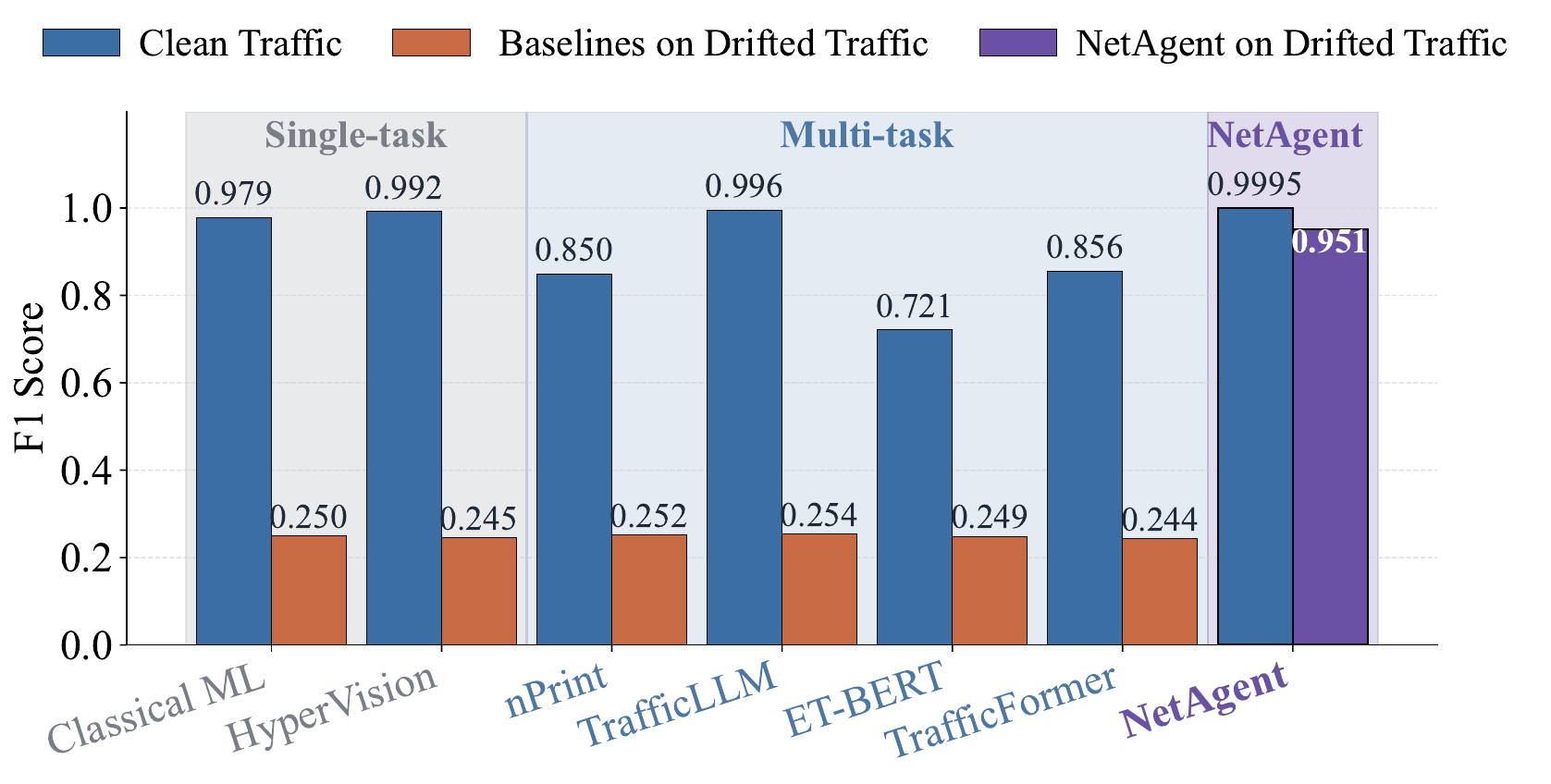}
\caption{Robustness to network drift. Baseline F1 drops to around 25\%, whereas \tool maintains 95.1\% F1 with only a 4.85-point decrease.}
\label{fig:noise}
\vspace{-2ex}
\end{figure}

As Figure~\ref{fig:noise} shows, all baselines collapse. The best single-task and multi-task baselines
%, \confirm{Classical ML} and \confirm{nPrint}, 
lose 74.7 and 74.2 F1 points, and all baselines fall to around 25\% F1. These methods are trained under a closed-world assumption in which every flow matches a known attack or benign pattern, so unseen background traffic is forced into the closest known pattern, producing many false positives. In contrast, \tool drops by only 4.85 points and achieves 95.1\% F1. Its planner retrieves relevant attack rules and flags only flows that exhibit the corresponding attack patterns. When classifier tools label background flows as attacks, the decision maker cross-checks those outputs against rule matches and traffic statistics to filter most false positives.

% \begin{table}[t]
%   \centering
%   \caption{Robustness to data drift on CIC-IDS2017: 20\% of the test set is
%   replaced with real backbone traffic (MAWI/WIDE, June 6, 2026, over 283M
%   packets).}
%   \label{tab:drift}
%   \scriptsize
%   \begin{tabular*}{\columnwidth}{@{}l@{\extracolsep{\fill}}ccc@{}}
%     \toprule
%     \textbf{Method} & \textbf{F1 (clean)} & \textbf{F1 (drift)} & \textbf{Drop} \\
%     \midrule
%     TrafficLLM    & 0.9955 & 0.2537 & $-74.52\%$ \\
%     ET-BERT       & 0.9990 & 0.2488 & $-75.10\%$ \\
%     TrafficFormer & 0.9985 & 0.2440 & $-75.56\%$ \\
%     nPrint        & 0.9980 & 0.2515 & $-74.80\%$ \\
%     IDS-Agent     & 0.9890 & 0.2452 & $-75.21\%$ \\
%     Classical ML  & 0.9960 & 0.2502 & $-74.88\%$ \\
%     \midrule
%     \textbf{\tool} & \textbf{0.9995} & \textbf{0.9510} & $-4.85\%$ \\
%     \bottomrule
%   \end{tabular*}
% \end{table}

Robustness also depends on how each component handles imperfect inputs. For the planner, we issue 100 single-task queries for web attack detection and 100 multi-task queries, such as jointly analyzing DDoS attacks and DGA domains. It produces valid workflows for 98 and 96 queries, respectively, only a 2-point drop for more ambiguous multi-task requests. For the executor, 88\% of generated snippets run successfully on the first attempt. The remaining failures are 8\% tool-parameter type errors, 3\% undefined references, and 1\% import errors, all implementation-level rather than logical workflow errors; one repair iteration raises success above 99\%. For the decision maker, classifier outputs may conflict. Across the four tasks in Figure~\ref{fig:robust-adaptive}(b), intermediate conclusions differ by up to 45.51 F1 points. Rather than choosing the best classifier or majority vote, the decision maker treats disagreement as a signal for further analysis and reasons over all results as complementary evidence. Its final conclusion exceeds the best individual conclusion by 2.26 F1 points on average.

\begin{takeaway}
    \textbf{Takeaway:} Under real-world background traffic, all baselines collapse because closed-world training gives them no notion of unfamiliar traffic, whereas \tool's performance stays robust.
\end{takeaway}

%%%%%%%%%%%%%%%%%%%%%%%

\subsection{RQ4: Ablation Study}
\label{sec:abl}

\begin{figure}[t]
\centering
\includegraphics[width=\columnwidth]{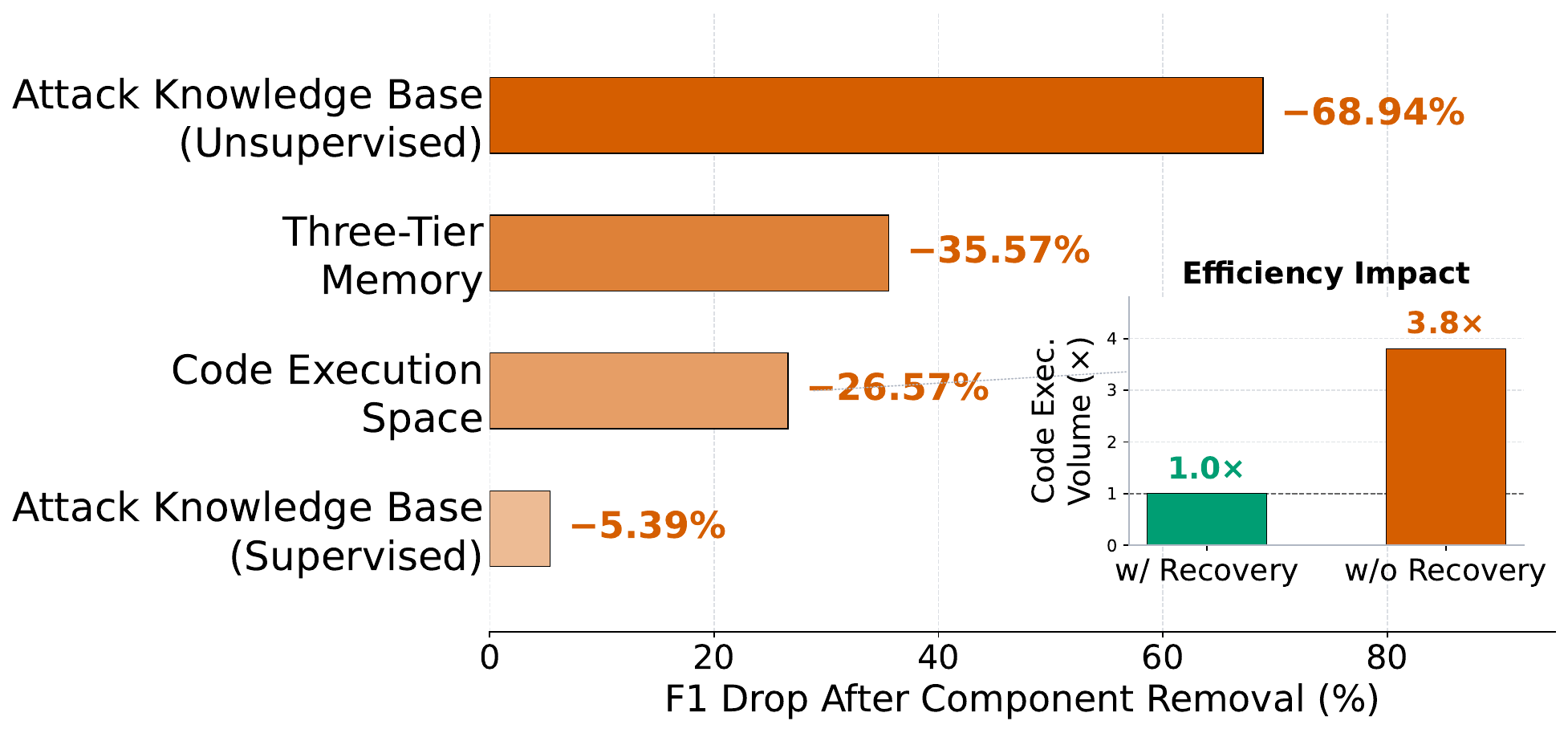}
\caption{Ablation on web attack detection. We remove individual \tool components and measure their impact on F1.}
\label{tab:ablation}
\vspace{0ex}
\end{figure}

We evaluate the contribution of each component by removing one at a time on the web attack detection task. As shown in Figure~\ref{tab:ablation}, replacing the three-tier memory with flat natural-language notes causes the largest drop, 35.57 F1 points, because the decision maker can no longer reconcile conflicting conclusions across iterations. Replacing the code execution space with ReAct-style tool calls over the same library reduces F1 by 26.57 points due to limited flexibility in executing complex workflows. Removing the attack knowledge base reduces F1 by 5.39 points when classifier tools remain available, but by 68.94 points in the unsupervised setting due to the loss of domain knowledge. Removing failure recovery increases executed code volume by 3.8$\times$ because of excessive re-execution, with an F1 change of 10.05 point.

The LLM drives planning, code generation, and decision making. We hence study sensitivity on web attack detection by replacing one agent's LLM at a time among GPT-5.5 (default), DeepSeek-V4-Flash (284B), and Qwen3.5-9B. As shown in Table~\ref{tab:llm}, using GPT-5.5 for all three agents performs best overall, with DeepSeek-V4-Flash a close second within 0.02 F1 points. Replacing the executor's LLM with Qwen3.5-9B increases generated code by 4.9$\times$ but reduces F1 by only 0.06 points, as the code execution space and failure recovery correct additional errors. Replacing the planner shortens workflows to 80\% of their original length and reduces F1 by 8.22 points, and replacing the decision maker reduces F1 by 17.54 points. Thus, the planner and decision maker benefit more from stronger LLMs, while the executor is less sensitive to model choice.

\begin{takeaway}
	\textbf{Takeaway:}  Every component contributes to performance. The planner and decision maker are more sensitive to LLM choice, whereas the executor is less sensitive because the code execution space and failure recovery correct errors from weaker models.
\end{takeaway}

% \begin{table}[!ht]
% \centering
% \caption{Ablation on CIC-IDS2017: effect of removing one component from the full \tool on the affected metric.}
% \label{tab:ablation}
% \scriptsize
% \setlength{\tabcolsep}{2pt}
% \begin{tabular*}{\columnwidth}{@{\extracolsep{\fill}}p{2.6cm}p{2.6cm}p{2.2cm}}
% \toprule
% \textbf{Component removed} & \textbf{Metric affected} & \textbf{Measured effect} \\
% \midrule
% Three-level memory $\to$ file-based (Section~4.5) & F1 & $\downarrow$35.57\% \\
% Code tool space $\to$ tool calls (Section~4.4) & F1 & $\downarrow$26.57\% \\
% Augmented knowledge retrieval (Section~4.3) & F1 & $\downarrow$5.39\% \\
% Augmented knowledge retrieval (unsupervised) & F1 & $\downarrow$68.94\% \\
% Fast failure recovery (Section~4.6) & code exec. volume & $\times$3.8 \\
% \bottomrule
% \end{tabular*}
% \end{table}

\begin{table}[t]
\centering
\caption{Impact of LLM selection on \tool performance.}
\label{tab:llm}

\scriptsize
\setlength{\tabcolsep}{2pt}
\renewcommand{\arraystretch}{1.15}

\begin{tabularx}{\columnwidth}{
    @{}
    >{\centering\arraybackslash}X
    >{\centering\arraybackslash}X
    >{\centering\arraybackslash}X
    >{\centering\arraybackslash}p{0.85cm}
    >{\centering\arraybackslash}p{0.85cm}
    >{\centering\arraybackslash}p{0.85cm}
    @{}
}
\toprule
\textbf{Planner}
& \textbf{Action}
& \textbf{Decision}
& \textbf{F1}
& \makecell{\textbf{Code}\\\textbf{Exec.}}
& \makecell{\textbf{Plan}\\\textbf{Steps}}
\\
\midrule

DeepSeek
& DeepSeek
& DeepSeek
& 0.9993
& 1.1$\times$
& 1.2$\times$
\\

\rowcolor{gray!7}
Qwen3-9B
& Qwen3.5-9B
& Qwen3-9B
& 0.7745
& 5.2$\times$
& 0.7$\times$
\\

GPT-5.5
& GPT-5.5
& DeepSeek
& 0.9994
& 1.0$\times$
& 1.0$\times$
\\

\rowcolor{gray!7}
Qwen3.5-9B
& GPT-5.5
& GPT-5.5
& 0.9173
& 3.7$\times$
& 0.8$\times$
\\

GPT-5.5
& Qwen3.5-9B
& GPT-5.5
& 0.9989
& 4.9$\times$
& 1.1$\times$
\\

\rowcolor{gray!7}
GPT-5.5
& GPT-5.5
& Qwen3.5-9B
& 0.8241
& 1.1$\times$
& 1.2$\times$
\\

\midrule

\textbf{GPT-5.5}
& \textbf{GPT-5.5}
& \textbf{GPT-5.5}
& \textbf{0.9995}
& 1.0$\times$
& 1.0$\times$
\\

\bottomrule
\end{tabularx}
\vspace{-0ex}
\end{table}

%% file: sections/discussion.tex
\section{Discussion}
%-------------------------------------------------------------------------------

\tool is not a real-time traffic processor. Each agent-loop iteration invokes LLMs for planning, code generation, and decision making, and the executor may run multiple analyses over a capture. Its latency is therefore dominated by iterative reasoning rather than packet arrival rate. This is a deliberate tradeoff. \tool answers open-ended analyst queries that require reasoning over intermediate evidence, whereas fixed detectors answer predefined questions quickly but cannot adapt to new ones. The two are complementary. \tool also supports converting confirmed findings into detection or filtering rules that can be enforced in real time by existing IDSs and firewalls.

\tool's capability is also bounded by its tool library and attack knowledge base. Tasks without a matching tool fall back to LLM-generated code, and attacks absent from the knowledge base must be analyzed using other available tools, such as classifiers or anomaly detectors. This is a coverage limitation rather than an architectural one. However, \tool's current tool space already includes 150+ verified tools covering diverse traffic analysis tasks. \tool also allows new tools and knowledge to be added easily, expanding its capabilities without changing the rest of the framework. Its goal is not to replace strong task-specific methods, but to turn them into composable building blocks for unified multi-task analysis, with capabilities that grow as more methods are added.

\tool currently adopts a hierarchical, role-based architecture, with the decision maker coordinating the planner and executor. Alternative designs include peer-to-peer agents, fixed role pipelines, and more decentralized multi-agent systems~\cite{camel,metagpt,autogen}, which can provide greater parallelism or reasoning diversity at the cost of additional communication and coordination. We choose this hierarchical design so the decision maker can maintain a global view and coordinate planning, execution, and memory across iterations. A possible extension is to assign hosts, time windows, or attack hypotheses to parallel planner-executor workers and consolidate their findings at a higher-level decision maker.

%% file: sections/conclusion.tex
%-------------------------------------------------------------------------------
\section{Conclusion}
%-------------------------------------------------------------------------------

We presented \tool, the first unified agentic framework for multi-task network traffic analysis. \tool combines knowledge-augmented planning, a comprehensive tool action space, code-based execution, failure recovery, and three-tier memory in an agent loop. Across diverse traffic analysis tasks, \tool achieves strong performance, generalizes to unseen distributions, and remains robust under realistic background noise, significantly outperforming existing methods. Our results show that agentic design can unify task-specific methods into a high-performing multi-task framework. Future work includes real-time integration, broader tool and knowledge coverage, and more parallel multi-agent architectures.

%% file: sections/app_toolkit.tex
% \section{Ethical Considerations}
% \label{appendix:ethics}

% \tool is evaluated exclusively on public research datasets and offline traffic captures. We do not collect private or personally identifiable traffic, monitor live networks, or interact with real-world network infrastructure.

\section{Additional Experimental Details}
\label{app:results}

\begin{table}[!ht]
\centering
\caption{Configuration settings of \tool.}
\label{tab:hyperparameters}
\scriptsize
\setlength{\tabcolsep}{6pt}
\renewcommand{\arraystretch}{1.15}

\begin{tabularx}{\columnwidth}{@{}>{\raggedright\arraybackslash}X c@{}}
\toprule
\textbf{Parameter} & \textbf{Value} \\
\midrule
Retrieved rules in retrieval-augmented planning ($k$) 
& 10 \\

Retrieved memories for decision maker 
& 30 \\

Maximum number of $\mathcal{E}$ records 
& $20{,}000$ \\

Maximum number of $\Sigma$ records 
& $500$ \\

Maximum number of $\Pi$ records 
& $50$ \\

Novelty weight ($\alpha$) 
& 0.6 \\

Timeliness threshold ($\tau_{\mathrm{age}}$) 
& 60 \\

Frequency threshold ($\tau_{\mathrm{freq}}$) 
& 0.1 \\

Conflict threshold ($\tau_{\mathrm{conflict}}$) 
& 0.5 \\

Maximum sandbox repair iterations 
& 4 \\

Maximum agent loops 
& 30 \\

\bottomrule
\end{tabularx}

\end{table}

Table~\ref{tab:hyperparameters} summarizes the configuration settings of \tool.
Table~\ref{tab:extraction_ledger_9tasks} and Table~\ref{tab:tools} present representative tools implemented in \tool, which constitute its tool action space.
Table~\ref{tab:detailed_results_full} provides additional results for \tool and the baseline approaches across evaluation tasks.

\begin{table*}[htbp]
\centering
\caption{Representative tools implemented and verification results.}
\label{tab:extraction_ledger_9tasks}
% 使用 resizebox 适应页面宽度
\resizebox{0.95\textwidth}{!}{%
\begin{tabular}{l c l l l}
\toprule
\textbf{Traffic Analysis Domains} & \textbf{\# Tools} & \textbf{Representative Tools} & \textbf{Dataset} & \textbf{Paper $\rightarrow$ Reproduced} \\
\midrule
Network Intrusion Detection & 22 & ML classifiers, flow anomaly scoring & UNSW-NB15 & F1: 0.9780$\to$0.9758 \\
Web Attack Detection & 10 & Robust CLFs, IOC lookup & CIC-IDS2017 & F1: 0.994 $\to$ 0.9995 \\
Fine-grained Attack Classification & 16 & Shallow learners, feature selection & CIC-IDS2018 & F1: 0.9613$\to$0.9590 \\
DDoS Attack Detection & 12 & Sketch detector, graph analysis & CIC-DDoS2019  & F1: 0.9992$\to$0.9988  \\
IoT Intrusion \& Device ID & 14 & IoT device ID, event fingerprinting & CICIoT2023 & F1: 0.7500$\to$0.7470 \\
VPN / Encrypted Traffic Classification & 20 & TLS encoder, Handshake FP & ISCX-VPN2016 & F1: 0.9890$\to$0.9883  \\
Website Fingerprinting & 15 & WF classifiers, multi-tab attacks & DF2018 & F1: 0.9830$\to$0.9822 \\
BGP Anomaly Detection & 9 & AS validator, semantics-aware hijack & CAIDA AS & F1: 1$\to$1 \\
DGA Malicious Domain Detection & 14 & DGA detector, DNS tunneling & DGArchive & F1: 0.9900$\to$0.9904 \\
\midrule
Infrastructure \& Utilities & 25 & PCAP parsing, traffic generation & N/A / Multiple & - \\
\midrule
\textbf{Total} & \textbf{157} & & & \\
\bottomrule
\end{tabular}%
}
\end{table*}

\begin{table*}[!t]
\centering
\caption{Detailed list of tools across traffic analysis sub-domains, extracted from 50+ published systems.}
\label{tab:tools}
\footnotesize
\renewcommand{\arraystretch}{1.2}
% 调整列宽比例以适应更长的子领域名称，总宽度0.95留出合适的边距和列间距
\resizebox{0.95\textwidth}{!}{%
\begin{tabular}{p{0.06\textwidth} p{0.22\textwidth} p{0.34\textwidth} p{0.33\textwidth}}
\toprule
\textbf{\# Tools} & \textbf{Sub-domain} & \textbf{Representative Tools} & \textbf{Source} \\
\midrule
% --- 基础解析层 ---
8 & \textbf{Traffic Parsing \& Preprocessing} & PCAP parsing, flow splitting, protocol identifier, traffic summary & nPrint~\cite{nprint}, netUnicorn~\cite{beltiukov2023search} \\
% --- 通用与Agent检测层 ---
12 & \textbf{General NIDS \& Anomaly Scoring} & Flow anomaly scoring, noise-robust detection, point-cloud analysis & PointCloud~\cite{pointcloud}, RAPIER~\cite{rapier}, HyperVision~\cite{hypervision} \\
10 & \textbf{Agent-based \& LLM-driven IDS} & LLM agent orchestration, BLE embedded IDS, rule generation & IDS-Agent~\cite{idsagent}, OASIS~\cite{cayre2024oasis}, RulingUnruly~\cite{ruling} \\
% --- 特定攻击类型层 ---
10 & \textbf{Web \& App-Layer Attack Detection} & Point-cloud false-positive reduction, robust encrypted malicious traffic detection & PointCloud~\cite{pointcloud}, RAPIER~\cite{rapier}, HyperVision~\cite{hypervision}, BadDesignSmells~\cite{flood2024bad} \\
8 & \textbf{DDoS \& Volumetric Attacks} & Routing-oblivious per-flow measurement, carpet-bombing detection & Lemon~\cite{lemon}, NetRadar~\cite{netradar} \\
4 & \textbf{Tunneling \& Stealthy Attacks} & Interest flooding analysis, tunneled flooding detection & GNN4IFA~\cite{agiollo2023gnn4ifa}, TunneledFlooding~\cite{fu2024detecting} \\
16 & \textbf{Fine-grained Attack Classification} & Temporal fusion encoding, transfer-robust feature learning, representation learning analysis & TFE-GNN~\cite{tfegnn}, Swallow~\cite{swallow}, Sugar~\cite{sugar}, CapsuleFormer~\cite{zhou2024capsuleformer} \\
% --- IoT与边缘安全层 ---
8 & \textbf{IoT Device Fingerprinting} & ML-based device identification, packet-level sequence matching & IoT-DeviceID~\cite{iotdeviceid}, WiFinger~\cite{wifinger} \\
6 & \textbf{IoT Malware \& C2 Detection} & C2 server revelation, allowlist-based detection & C2Mine~\cite{davanian2024c2miner}\\
% --- 加密与匿名流量层 ---
10 & \textbf{VPN \& Proxy Classification} & TCP-aware traffic augmentation, encapsulated TLS fingerprinting, cross-layer RTTs & Rosetta~\cite{rosetta}, ProxyFP~\cite{proxyfp}, CrossRTT~\cite{crossrtt}, GFW-Encrypted~\cite{gfwencrypted} \\
10 & \textbf{Encrypted Traffic Representation} & Pre-trained datagram representation, multi-scale contrast, traffic refinery & ET-BERT~\cite{etbert}, TFE-GNN~\cite{tfegnn}, MultiScaleContrast~\cite{panda2024multi}, TrafficRefinery~\cite{bronzino2021traffic} \\
15 & \textbf{Website Fingerprinting (Offense \& Defense)} & Multi-tab attacks, fine-grained at scale, MI minimization defense, RTT-aware transduction & MultiTab-WF~\cite{multitabwf}, Oscar~\cite{oscar}, Swallow~\cite{swallow}, FRUGAL~\cite{frugal}, CELLSHIFT~\cite{jansen2026cellshift}, AVXProbe~\cite{kim2025avxprobe}, RealisticWF~\cite{jeon2026rofire}, StopDontClick~\cite{mitseva2024stop}, TransformerWF~\cite{zhou2023wf} \\
% --- 特定协议与场景层 ---
14 & \textbf{Domain \& DNS Security} & Robust DGA classification, large-scale cryptomining traffic detection & Joint-Adv-DGA~\cite{robustdga}, MineShark~\cite{mineshark} \\
9 & \textbf{Routing \& Topology Anomaly} & Semantics-aware routing anomaly detection, open-world normality shift adaptation & Semantics-Aware~\cite{semanticsaware}\\
% --- 基础模型与辅助工具层 ---
12 & \textbf{Traffic Foundation Models} & Efficient pre-training, LLM generic representation, multi-modal pre-training & TrafficFormer~\cite{trafficformer}, TrafficLLM~\cite{trafficllm}, MM4flow~\cite{mm4flow} \\
6 & \textbf{AutoML \& Automated Analysis} & Automated traffic analysis, LLM agent orchestration & nPrint~\cite{nprint}, netUnicorn~\cite{beltiukov2023search} \\
7 & \textbf{Data Augmentation \& Robustness} & Network data augmentation, certified robustness, concept drift navigation & NetDiffusion~\cite{netdiffusion}, CertTA~\cite{certta}, LEAF~\cite{leaf}, BARS~\cite{wang2023bars}, PANTS~\cite{jin2025robustifying} \\
\midrule
\multicolumn{1}{r}{\textbf{157}} & \multicolumn{1}{l}{\textbf{Total}} & \multicolumn{2}{l}{\textit{Spanning 50+ published systems across 17 fine-grained sub-domains}} \\
\bottomrule
\end{tabular}
}
\end{table*}

\begin{table*}[!t]
\centering
\caption{Detailed F1 performance of \tool and all 28 baselines across 9 benchmarks.}
\label{tab:detailed_results_full}
\footnotesize
\resizebox{0.85\textwidth}{!}{%
\begin{tabular}{@{} l c c c c c c c c c @{}}
\toprule
\textbf{Method (Type)} & \textbf{NIDS} & \textbf{Web} & \textbf{Fine} & \textbf{IoT} & \textbf{DDoS} & \textbf{VPN} & \textbf{WF} & \textbf{BGP} & \textbf{DGA} \\
\midrule
\multicolumn{10}{l}{\textbf{Proposed Method}} \\
\textbf{\tool (Ours)} & \textbf{0.9863} & \textbf{0.9995} & \textbf{0.9999} & \textbf{0.9998} & \textbf{0.9999} & \textbf{0.9242} & \textbf{0.9965} & \textbf{1.00} & \textbf{0.9999} \\
\midrule
\multicolumn{10}{l}{\textbf{Multi-task Baselines (5)}} \\
nPrint & 0.9807 & 0.8499 & 0.8714 & 0.8232 & 0.9850 & 0.5478 & 0.7503 & 0.5948 & 0.9750 \\
TrafficFormer & 0.8144 & 0.8558 & 0.9985 & 0.7896 & 0.8128 & 0.8441 & 0.9520 & 0.5194 & 0.3556 \\
ET-BERT & 0.6556 & 0.7209 & 0.8443 & 0.6099 & 0.7545 & 0.9454 & 0.6898 & 0.6073 & 0.2588 \\
TrafficLLM & 0.7443 & 0.9955 & 0.8191 & 0.7125 & 0.7876 & 0.8271 & 0.7239 & 0.5504 & 0.5257 \\
ReAct-Agent & 0.7187 & 0.7505 & 0.7233 & 0.8262 & 0.7556 & 0.3967 & 0.4997& 0.9752 & 0.4228 \\
\midrule
\multicolumn{10}{l}{\textbf{Classical ML Baselines (7)}} \\
Random Forest & 0.9663 & 0.9795 & 0.9995 & 0.9398 & 0.8675 & 0.8276 & 0.7503 & 0.4542 & 0.9899 \\
Gradient Boosting & 0.9722 & 0.9680 & 0.9999 & 0.9199 & 0.9611 & 0.7327 & 0.7447 & 0.4291 & 0.9556 \\
KNN & 0.9652 & 0.9553 & 0.9486 & 0.9197 & 0.9193 & 0.7182 & 0.6246 & 0.5824 & 0.8419 \\
SVM & 0.9471 & 0.5945 & 0.9101 & 0.8996 & 0.8924 & 0.4724 & 0.6429 & 0.5817 & 0.8141 \\
Decision Tree & 0.9667 & 0.9695 & 0.8790 & 0.9380 & 0.8988 & 0.5543 & 0.6328 & 0.6971 & 0.9554 \\
Logistic Regression & 0.9198 & 0.6419 & 0.9223 & 0.9202 & 0.8157 & 0.7513 & 0.5101 & 0.4550 & 0.7615 \\
KMeans & 0.4995 & 0.5226 & 0.5307 & 0.6520 & 0.6324 & 0.5861 & 0.5842 & 0.4069 & 0.5544 \\
\midrule
\multicolumn{10}{l}{\textbf{Single-task SOTA Baselines (16)}} \\
HyperVision & — & 0.9920 & — & — & — & — & — & — & — \\
PointCloud & 0.9750 & — & — & — & — & — & — & — & — \\
IDS-Agent & — & — & — & 0.9691 & — & — & — & — & — \\
Rosetta & — & — & — & — & — & 0.9120 & — & — & — \\
DeepFP & — & — & — & — & — & — & 0.9797 & — & — \\
Semantics-Aware & — & — & — & — & — & — & — & 1.00 & — \\
RAPIER & — & 0.9849 & — & — & — & — & — & — & — \\
Lemon & — & — & — & — & 0.9989 & — & — & — & — \\
NetRadar & — & — & — & — & 0.9163 & — & — & — & — \\
TFE-GNN & — & — & 0.9976 & — & — & — & — & — & — \\
Sugar & — & — & 0.9941 & — & — & — & — & — & — \\
Swallow & — & — & — & — & — & — & 0.9692 & — & — \\
Multi-tab WF & — & — & — & — & — & — & 0.9576 & — & — \\
IoT Device ID & — & — & — & 0.8125 & — & — & — & — & — \\
WiFinger & — & — & — & 0.7448 & — & — & — & — & — \\
Joint Adversarial DGA & — & — & — & — & — & — & — & — & 0.9851 \\
\bottomrule
\end{tabular}%
}
\begin{flushleft}
\footnotesize
\textit{Note:} Task abbreviations: NIDS (UNSW-NB15), Web (CIC-IDS2017), Fine (CIC-IDS2018), IoT (CICIoT2023), DDoS (CIC-DDoS2019), VPN (ISCX-VPN2016), WF (DF2018), BGP (CAIDA), DGA (DGArchive).
\end{flushleft}
\end{table*}

%% file: sections/app_prompts.tex
\section{Prompts}
\label{app:prompts}

Below, we provide the priompt templates used by \tool for workflow planning, tool selection, code execution, and decision making. These prompts are dynamically populated with relevant context.

\subsection{Knowledge-Augmented Planning}
Before workflow generation, \tool retrieves attack rules relevant to the planning request from the attack knowledge base. The retrieved rules are incorporated into the planner prompt to provide domain knowledge for workflow generation and tool selection.

\begin{lstlisting}[style=flowprompt]
You are in the pre-planning data-reading phase. Your ONLY job is to read and retrieve raw data BEFORE the main analysis plan is created.

## Your Task
Given the user query, call tools that READ data from files or directories so the planner can see the actual layout. You may ONLY call:

- **File / directory operations** -- list_directory, list_directory_contents, find_files, file_info, check_is_directory, read_file
- **Table previews (bounded)** -- read_csv, read_csv_sample, read_large_csv, load_kdd99
- **pcap basic info ONLY** -- pcap_quick_statistics (O(1)-memory overview: protocol mix, top IPs/ports, time range)

You must NOT call any tool that parses a pcap into flows, extracts features, analyzes/classifies/detects, trains ML models, runs inference, transforms data, or generates traffic. In particular, DO NOT call: parse_pcap, fast_pcap_parser, parse_pcap_streaming, sample_flows_from_pcap, extract_flow_features, extract_flow_features_cicflow, extract_packet_length_sequence, packet_length_distribution, flow_splitter, flow_correlation, connection_graph_builder, merge_datasets, extract_dns_queries, protocol_identifier, bgp_update_reader, traffic_summary, clean_numeric_features -- these belong to the EXECUTION steps, not pre-plan.

## Output Format
Return a JSON object:
{{
  "rationale": "Brief explanation of what information you need",
  "tool_calls": [
    {{"tool": "tool_name", "params": {{"param1": "value1", ...}}}}
  ]
}}

If no information gathering is needed, return {{"tool_calls": []}}.

## Important
- Extract parameter values from the user query (e.g. file paths, URLs, IPs).
- If you need a file path but the query doesn't specify one, call list_directory / find_files first to discover files.
- Keep parameter names EXACTLY as shown in the tool specs.
- You may call multiple tools in one round, but keep it focused -- only what's needed to read the data.
- Previous round findings are shown below; use them to decide if more context is needed.

## LARGE-FILE POLICY (PCAP)
- For PCAP files use ONLY pcap_quick_statistics (O(1)-memory overview) -- NEVER a full parse
  or feature extraction in this phase. The system auto-caps pcap tools (max_packets), but
  always pass explicit limits: e.g. pcap_quick_statistics(max_packets=10000).
- If the file size is unknown, run file_info or list_directory first to learn total size,
  then decide. Do NOT parse a pcap just to learn its size.
- Keep gathering lightweight -- you only need enough context (file layout + basic pcap stats)
  for the planner to start. Full flow extraction, feature engineering, DNS/TLS metadata,
  and anomaly analysis belong to the EXECUTION steps (planner + FlowAct), not this phase.
\end{lstlisting}

\subsection{Workflow Generation}
The planner generates an execution workflow based on the planning request, observed traffic, available tools, and retrieved attack knowledge.

\begin{lstlisting}[style=flowprompt]
You are an expert specializing in developing implementation plans based on user queries.

## Your Goal
- Analyze the user query and output a step-by-step plan.
- Review your generated plan for potential issues and refine it (do not include this step in the output).
- Ensure the plan covers the user query (do not include this step in the output).

## Worked Example (use this structure as a reference; adapt content to the actual query)
### User query:
Extract network traffic features from file.pcap.
### YOUR RESPONSE:
{"plan":[
{"step":1,"action":"Parse the PCAP file `file.pcap` with the fast parser (dpkt engine, streams the file).","tools":["fast_pcap_parser"],"databases":[],"expected_inputs":["file.pcap"],"expected_outputs":["parsed_traffic"]},
{"step":2,"action":"Extract features from the parsed file.","tools":["extract_flow_features"],"databases":[],"expected_inputs":["step1_output.parsed_traffic"],"expected_outputs":["flow_features"]}
]}

## CRITICAL: Incremental Planning
- Do NOT generate a complete plan for the entire task. Generate only the FIRST 2-3 steps.
- The first step should be data loading / exploration / parsing.
- Subsequent steps will be determined after reviewing the results of these initial steps.
- Think of this as an iterative analysis: you start by looking at the data, then decide what to do next.

## Layered Analysis: Layer 0 = Preliminary Conclusion
- This task uses layered analysis. Your FIRST batch (2-3 steps) is Layer 0 and must:
  1. Load/explore the data using FAST scanning tools (see LARGE-FILE POLICY below).
  2. Extract key features from the loaded data (protocol_identifier / traffic_summary /
     extract_flow_features / extract_dns_queries on sampled flows).
  3. End with a step that synthesizes an explicit "Preliminary Conclusion" -- a plain-text
     assessment of the key features (packet/flow counts, protocol mix, top IPs/ports,
     time range, any anomalies). This conclusion is shown to the user immediately.
- Do NOT plan the whole workflow. Deeper refinement happens in later replan rounds.

## LARGE-FILE POLICY (PCAP)
- If the target is a PCAP/PCAPNG file that may be large (>~50-100 MB or unknown size):
  * Step 1: pcap_quick_statistics for an O(1)-memory overview (protocol mix, top IPs/ports, time range).
  * Flow extraction: fast_pcap_parser / extract_flow_features_cicflow -- both dpkt-based, stream the file,
    never load it fully, and their defaults are NATIVE (max_flows=0, sample_ratio=1.0, max_packets=0):
    they read ALL flows and full packet detail by default, so do NOT set max_flows / sample_ratio / seq_len.
  * parse_pcap()/parse_pcap_streaming() are dpkt-based (same engine as fast_pcap_parser) and stream the file,
    but store full per-packet raw detail by default -- bound max_packets (e.g. 10000-50000) to keep runs short.
  * For VERY LARGE captures (>~1GB), do NOT plan a single unbounded extraction call -- it runs for minutes with
    no checkpoint. Plan the extraction step to loop in batches INSIDE its code: call extract_flow_features_cicflow
    repeatedly with batch_packets=<N> (e.g. 2_000_000) and resume_from=<previous packets_consumed> until is_complete,
    collect each batch's flow_features, then combine with impl.merge_flow_features(batches). This keeps every call
    bounded and resumable, and prints live progress to stderr.
- For small files (<~50 MB) any parser is fine.

## PCAP + LABELED CSV (CICFlowMeter) CLASSIFICATION WORKFLOW
- When the user supplies a pcap AND a CICFlowMeter CSV (80+ flow features + a Label column) and wants to
  classify/detect malicious traffic FROM THE PCAP, use this chain (features come from the pcap; the CSV is
  only the ground-truth label source -- do NOT train on the CSV's feature columns):
  1) extract_flow_features_cicflow(pcap_path=...) -- dpkt streaming,
     O(#flows) memory, aggregates Fwd/Bwd CICFlowMeter-style features + TTL/flags/window + pkt_len_seq.
     NOTE: max_flows / seq_len / max_packets default to 0 = unlimited (native), so it reads ALL flows
     and the full pkt_len_seq by default. Do NOT set max_flows / seq_len unless you want to BOUND the
     output (e.g. a quick preview).
  2) build_flow_classification_dataset(flow_features=stepN_output['flow_features'], csv_path=...,
     label_col='Label') -- csv_path may be the labels FOLDER (all CSVs merged, covers CICIDS2017's
     split per-session label files) or a single CSV; aligns flows to label rows by 5-tuple
     (direction-insensitive), builds data_X / data_Y / binary_Y.
     Prefer passing the labels directory from the user's query (or a file found by pre-plan) over
     guessing a filename. If the folder mixes other days, pass file_prefix='Friday' (or the day being
     analyzed) so only that day's label files are merged.
  3) split_dataset(data_X=..., data_Y=..., test_size=0.2) -- stratified by default (preserves class
     proportions; important for imbalanced data).
  4) Run SEVERAL classifiers on the SAME split -- predict_ml_random_forest, predict_ml_svm,
     predict_ml_knn, predict_ml_gradient_boosting. Do NOT rely on a single classifier.
  5) COMPARE them: impl.compare_classification_results(results=[{name, y_true, y_pred}, ...]) ->
     best_method / agreement / majority-vote accuracy. Then evaluate_binary_classification on the
     chosen (or best) model -> accuracy/precision/recall/F1.
- If the label distribution is highly imbalanced (any class <~5%), pass class_weight='balanced' to
  predict_ml_random_forest / predict_ml_svm and keep the stratified split.
- If data_Y is multi-class, prefer build_flow_classification_dataset's binary_Y (0=benign, 1=attack);
  convert_multilabels_to_binary is the fallback for numeric labels.
- The pre-plan phase only SAMPLES a small fraction of flows for a quick preview. For WHOLE-dataset
  detection/classification, plan a re-extraction step with extract_flow_features_cicflow -- its defaults
  already read ALL flows and full pkt_len_seq natively (max_flows=0, sample_ratio=1.0, max_packets=0),
  so do NOT set max_flows / sample_ratio / seq_len. Then align labels / train / evaluate on the full set.
  Do NOT classify only the pre-plan sample.

## Method Comparison & Benchmarking
- When a stage has multiple viable methods, plan them as SIBLING steps that SHARE the same namespace input (e.g. the same `stepN_output['data_X']`, `stepN_output['flow_features']`, or `stepN_output['parsed_traffic']`), ONE method per step. Examples:
  * Feature extraction: `extract_flow_features_cicflow` vs `extract_packet_length_sequence` vs `nprint_feature_encoder` on the same flows.
  * Classifiers: `predict_ml_random_forest` vs `predict_ml_svm` vs `predict_ml_knn` vs `predict_ml_gradient_boosting` on the same train/test split.
- After the sibling method steps, add ONE COMPARISON step with `tools: []` whose action begins with "COMPARE": its code calls each candidate via `impl.<tool>(...)` on the shared input, then summarizes with `impl.compare_classification_results(results=[{name, y_true, y_pred}, ...])` (per-method metrics table, best_method, agreement_rate, majority-vote labels/accuracy).
- Temporal / multi-period analysis: when the query spans multiple time periods, files, or days (e.g. CICIDS Monday vs Friday, or several pcaps), plan ONE analysis step PER period through the same pipeline, then a step that calls `impl.aggregate_period_results(results={period: {metric: value}, ...})` to surface the trend (best/worst period, first->latest change).
- Comparisons reveal CONSENSUS (high agreement -> high-confidence conclusion) and DIVERGENCE (disagreement -> worth deeper investigation). Plan follow-up steps for divergences.

## Planning Requirements
- Steps must be atomic, executable, and ordered to respect data dependencies.
- Explicitly define data flow: each step's `expected_outputs` must be referenceable by subsequent steps' `expected_inputs`.
- If the query is trivial, return a single-step plan that directly addresses it.
- Focus on the `user query` semantics; tools and databases are means to an end, not the planning target.
- **TOOL-SELECTION RULES (critical -- execution fails if these are broken):**
  * Every name in `tools` MUST be an EXACT, character-for-character match of a tool name from the provided catalog. Do not invent, abbreviate, or alter tool names (no extra underscores, no camelCase->snake_case changes, no typos).
  * **ONE tool per step.** Each step must have exactly one tool in its `tools` list. If a task needs multiple tools, split it into multiple steps -- one tool per step.
  * If a task does not clearly match any cataloged tool, put an EMPTY `tools: []` for that step rather than guessing a name.
  * Verify parameter compatibility: ensure input types, formats, and semantics match the function's expectations before assigning a tool.
  * Do not list a tool you cannot find in the catalog.

## Format Requirements
- Output MUST be valid, minified JSON with no trailing commas, comments, or markdown fencing.
- The root object must contain exactly one key: `"plan"`, whose value is a list of step objects.
- Each step object MUST contain the following fields:
- `"step"`: integer, starting from 1 and incrementing by 1 sequentially.
- `"action"`: string, detailed instructions for this step.
- `"tools"`: list of strings (tool names from the provided catalog), or empty list `[]` if none.
- `"databases"`: list of strings (database names from the provided catalog), or empty list `[]` if none. Databases are CALLABLE external data sources (e.g. RDAP / ThreatFox / crt.sh); list a database on a step ONLY when that step needs to enrich IP/domain/IOC/certificate evidence by querying an external source -- the generated code will call it as `dbs.<name>(...)`. If the step does not query an external source, leave `databases` empty.
- `"expected_inputs"`: list of strings referencing prior outputs using format `step<N>_output.<output_name>` or empty list `[]` if none.
- `"expected_outputs"`: list of strings naming the outputs this step produces, using snake_case identifiers.
\end{lstlisting}

\subsection{Workflow Tool Selection}
The planner selects appropriate tools from the tool action space for each workflow based on the request context and available tool descriptions.

\begin{lstlisting}[style=flowprompt]
You are an expert selecting relevant tools and databases for network traffic analysis.

## Your Task:
- Analyze the user query and memory context.
- Select ONLY the tools that are directly needed for this specific task.
- Be selective: typically 3-8 tools are sufficient, not all 116 tools.

## Selection Criteria:
- Choose tools that match the analysis goal (e.g., malware detection, anomaly detection, etc.)
- For tasks involving inspecting, reading, writing, finding, copying, moving, or deleting files, select generic tools (read_file, write_file, list_directory, find_files, copy_file, move_file, delete_file, create_directory, file_info) in addition to any analysis tools.
- Avoid selecting tools for unrelated tasks.
- If the query is general, select core analysis tools only.

## Return format (JSON):
{
  "tools": ["tool_name_1", "tool_name_2", ...],
  "databases": ["db_name_1", "db_name_2", ...],
  "rationale": "Brief explanation of why each tool was selected."
}

User query:{query}

Memory context:{memory_context}

Pre-Plan Context (data already gathered):{pre_plan_context}

Databases names: [{database_names}]

Database descriptions:
{database_catalog}

Tools names: [{tool_names}]
Tool descriptions:
{tool_catalog}
\end{lstlisting}

\subsection{Code Execution}
Each workflow generates multiple Python code snippets for its steps, which are then executed to perform the corresponding analysis.

\begin{lstlisting}[style=flowprompt]
You are a Python expert responsible for writing code based on specific tasks. Your code must be highly efficient and accurate.

## TOOLS Provided
- The following tool is available from the `impl` module: {tool_names}
- Call as `impl.<tool_name>(<param>=<value>, ...)`. Do NOT import the tool; `impl` is already in scope.
- Each tool returns a dict (typically containing `text_summary` and a task-specific result field). Pass that return value directly as the `result` argument to `record_tool`.
- **REQUIRED: call the tool and record its result.** Your code MUST call the tool and immediately pass its return to `record_tool`:
    result = impl.<tool_name>(required_param=<value>, optional_param=<value>)
    record_tool(step=<step_index>, result=result, tool_name="<tool_name>")
Do NOT print the result instead of recording it. Do NOT skip calling the tool.

### Tool specs:
{tool_docs}

## TOOL USAGE
- No specialized tool is available for this step. You MUST write custom code to complete it.
- Write COMPLETE, executable Python: never leave an empty or unfinished assignment (e.g. `x = ` with nothing after it), and define every variable before you use it. Compute derived values step by step with real code.
- At the end of the code, call `record_tool(step=<step_index>, result=<dict>, tool_name="step_logic")` EXACTLY ONCE. The `result` dict MUST include a `text_summary` key with a concise string description.

- You MAY call SEVERAL `impl.<tool>(...)` functions in this ONE snippet on the same input variables (e.g. predict_ml_random_forest, predict_ml_svm, predict_ml_knn, ... on the same stepN_output['data_X']).
- For classifiers: gather each method's y_pred, then call `impl.compare_classification_results(results=[{'name': 'RF', 'y_true': y_true, 'y_pred': pred_rf}, ...])` to produce the per-method metrics table, best_method, agreement_rate and majority vote.
- For time-period trends: call `impl.aggregate_period_results(results={'<period>': {'<metric>': value}, ...})`.
- Record EXACTLY ONE dict via `record_tool(step=<step_index>, result=result, tool_name="compare_methods")` that contains the comparison output (comparison_table / best_method / agreement_rate / majority_vote_labels / vote_accuracy) plus a `text_summary`. No logic after record_tool.

## Mandatory Requirements
- **TOOL-FIRST (highest priority):** if a provided tool can fulfill the step, CALL it via `impl.<tool_name>(...)`. Prefer delegating to a tool over writing equivalent custom code. Write custom code ONLY for what tools cannot cover.
- **MUST call the tool:** this step has a tool assigned. Your code MUST call `impl.<tool_name>(...)` with the correct parameters. Not calling the tool is a failure -- the step exists specifically to use this tool.
- **Keep it short:** call the tool, record the result, stop. No extra logic, no loops, no post-processing, no printing. If you need data from the result, the next step will use `stepN_output.<field>`.
- Execute ONLY the current step. Do not plan or write code for subsequent steps.
- **Record exactly once:** call `record_tool(step=<step_index>, result=<dict>, tool_name="<tool_name>")` exactly once per step. Never call it inside a loop, and never omit it. `record_tool` is built-in -- do NOT import it.
- **`record_tool` must be the LAST line.** Any code after `record_tool` will be ignored and is a waste of tokens. Do not add post-processing, printing, or variable extraction after recording.
- **Tool result is the record result:** pass the tool's return value dict directly as `result` to `record_tool`. Do not wrap it, do not replace it with a print statement.
- **Fallback (no tool applies):** if no tool can do the step, write custom code and pass a dict with at least a `text_summary` key as `result`. Use `tool_name="step_logic"` in this case.
- **`text_summary` is mandatory:** the `result` dict MUST always include a `text_summary` key with a short string value. Never omit it.
- **Serialization:** every key and nested value inside `result` must be msgpack-serializable (None, bool, int, float, str, bytes, list, dict). Convert numpy/pandas objects with `.tolist()` or `.item()` before recording.
- **No fabricated data:** never fabricate, guess, or falsify results. If a tool fails or data is missing, report the failure honestly (e.g., `text_summary: "tool failed: <reason>"`) rather than inventing output.
- **Parameter types:** carefully match the semantic meaning and type of each tool parameter. If the available data does not match a tool's expected type/format, add an explicit preprocessing/transform step rather than passing it through.
- **Pass tool inputs through, do not reconstruct:** if a tool's input is available as a variable from a previous step's output, pass it DIRECTLY -- e.g. `result = impl.flow_anomaly_scorer_ml(flow_features=step2_output['flow_features'])`. NEVER rebuild a list/dict by looping and overwriting a single variable, and NEVER manually reorder or re-type fields. Hand-reconstruction is a common source of silent bugs (overwritten loops, wrong container shape).
- **Context hygiene:** only reference variables that actually exist in the provided context. Do not modify variables unrelated to the current step.
- `dbs.<name>(...)` / `query_database("<name>", ...)` return a JSON-safe dict and never raise -- check the `error` field if present.
- **Code block format:** output your code inside a fenced code block (triple backticks ```). The code block must contain ONLY valid Python code. Never put natural language, bullets, or arrows inside the code block.

Workflow purpose: {workflow_purpose}
Current step index: {step_index}
Current step requirements: {step_action}
Current step expected input: {expected_inputs}
Current step expected output: {expected_outputs}
Available variables types: {context_snapshot}
Available variables meaning: {var_summary}
Memory context: {memory_context}

YOUR FIRST MOVE: read the Tools/Databases section of the system prompt and ask 'which provided tool does this step?'. If a tool matches, call it via `impl.<tool_name>(...)` and record its return value. Only write custom code for what the tool cannot cover.
\end{lstlisting}

\subsection{Decision Making}
Given the user's query, the current workflow, and the latest execution results, the decision maker either produces the final answer to the user's query or issues a new request for further analysis, triggering another iteration.

\begin{lstlisting}[style=flowprompt]
You are the DECISION MAKER of a network-traffic analysis agent. Your job is to translate the user's request into ONE concrete, actionable execution GOAL for the Planner (which then plans tool steps and the Executor runs them via code).

## What a good goal contains
- The exact data to operate on (the discovered data sources below, or the file/folder the user named).
- The analysis objective (e.g. load -> preprocess -> feature extraction -> classify -> evaluate metrics).
- Explicit success criteria (e.g. report F1/accuracy on a held-out test set; compare >=2 classifiers).
- Verification hints if the task is complex (e.g. 'verify the label alignment before training').

## Constraints
- Be SPECIFIC about data sources: if the query names a pcap/folder/csv, the goal must reference it and state whether to use the unified preprocessing (raw traffic field) first.
- A single goal, written in markdown. The Planner will turn it into steps.

Return JSON ONLY:
{
  "goal": "<markdown goal>",
  "success_criteria": ["..."],
  "data_source_hint": "<which input source to load first>"
}

You are the DECISION MAKER of a network-traffic analysis agent. You review the execution results of the current goal and decide the next action.

## Two responsibilities
1. JUDGE the current results: are they correct / complete / trustworthy? Is any verification or deeper analysis needed (e.g. a suspicious metric, an unverified label alignment, a scope problem, a single-method conclusion)?
2. DECIDE the next action: either CONTINUE with a refined/new goal for the Planner, or COMPLETE when the user's request is well satisfied.

##  Output contract (JSON only)
{
  "action": "continue" | "complete",
  "new_goal": "<markdown goal for the next plan round, empty if action==complete>",
  "issues": ["<problems found in the current results, empty if none>"],
  "verification_needs": ["<further verification/analysis needed, empty if none>"],
  "reason": "<one-paragraph rationale>"
}

## Rules
- Do NOT declare COMPLETE when metrics the user asked for (accuracy/F1) have not been computed, or when a classification conclusion relies on a single method -- the deterministic guard will override you anyway; be correct instead.
- COMPLETE only when the goal is satisfied with concrete evidence from the executed steps.
- When CONTINUE, give the Planner a SHARP new goal that fixes the gap (missing data, wrong scope, needs a compare step, needs verification of a specific artifact).
- ANALYZE, DON'T RESTATE: judge the numbers, don't repeat artifact summaries.

If you are the dynamic replanning controller for a network traffic analysis agent.

## Your Role
- Analyze the intent behind user queries.
- Based on executed step observations and accumulated analysis findings, decide whether more executable steps are needed.
- Return exactly one JSON object with schema:
{"need_more_steps": boolean, "reason": string, "new_steps": [{"step": step index:int, "action": specific action to take:str, "tools": the list of tools to use:list[str], "databases": the list of databases to use:list[str], "expected_inputs": the expected inputs for this step:list[str], "expected_outputs": the expected outputs for this step:list[str]}]}

## Blocking Issues to Resolve FIRST
{blocking_issues}
- If any blocking issue is flagged above, your FIRST new steps MUST resolve it (e.g. re-run extraction on the correct file, apply the day/file_prefix filter, verify which label CSV files were actually merged, or re-align to the right scope) BEFORE continuing model training/evaluation.
- Common case: a query asks for classification metrics (accuracy/F1/confusion) but no evaluate_* result exists yet -- add split_dataset -> predict_ml_* -> evaluate_binary_classification steps to produce them.
- Do NOT add split/train/evaluate steps on data you believe is out of scope while a blocking issue is unresolved.
- Only report need_more_steps=false when the query is genuinely answered AND no blocking issue remains.

## Incremental Analysis Approach
- You are building analysis incrementally, like a human analyst: each round of analysis builds on previous findings.
- Review the accumulated observations below -- they represent what has been learned so far.
- Based on these observations, decide what to investigate next.
- Generate only 1-3 new steps that dig deeper into the most promising findings.
- If a step failed, consider an alternative approach rather than retrying the same thing.

## MUST CONTINUE on failure
- If the most recent round contains a FAILED tool or empty result (e.g. "Insufficient...", "Error", "No data",
  an ML scorer that could not run), and the user's query is NOT yet fully answered, you MUST add replacement
  steps that achieve the same goal with OTHER allowed tools.
- Do NOT set need_more_steps=false just because a step failed or a tool was unusable. Only stop when the query is
  actually answered, or every reasonable available tool has been tried.
- Example: if flow_anomaly_scorer_ml fails, fall back to flow_anomaly_scorer_stat, ddos_detection,
  beaconing_detection, dns_tunnel_detection, or signature_matching -- do not stop the analysis.
- When switching tools, reuse variables already in the namespace (e.g. the same stepN_output['flow_features'])
  instead of re-parsing the whole file.

## Layered Refinement
- You are now planning Layer {layer_index} (a REFINED layer). The previous layer concluded:
{latest_conclusions}
- Generate 1-3 steps that DEEPEN the analysis: verify/refute earlier hypotheses with more
  specific evidence, drill into the most promising findings (top flows, suspicious IPs/ports,
  DNS/TLS metadata), and quantify.
- End this batch with one step that synthesizes this layer's "Refined Conclusion" (plain text).
- LARGE-FILE POLICY: prefer pcap_quick_statistics / fast_pcap_parser(sample_ratio) /
  sample_flows_from_pcap; do NOT re-parse the whole PCAP with parse_pcap on every layer.
  Reuse in-memory namespace variables from earlier steps when possible.
- If a PCAP+CSV classification chain needs more depth, add steps that deepen on the already-aligned
  dataset (e.g. more classifiers / a comparison step) or that re-run extract_flow_features_cicflow --
  its defaults already read ALL flows natively (max_flows=0), so do NOT set max_flows / sample_ratio.
  Do not re-parse the whole file on every layer; reuse in-memory namespace variables.

## Prefer Comparison over Single-Method Continuation
- To DEEPEN an analysis, prefer planning ALTERNATIVE methods on the SAME data and comparing them (a COMPARE step calling `impl.compare_classification_results(...)`) over re-running one method or stopping. This is how the best classifier / feature set is selected.
- For classification, ALWAYS compare multiple classifiers (predict_ml_random_forest / predict_ml_svm / predict_ml_knn / predict_ml_gradient_boosting) on the same train/test split with a COMPARE step -- do NOT conclude on a single classifier.
- If the label distribution is imbalanced (a class <~5%), use class_weight='balanced' on the supported classifiers and keep the stratified split.
- Temporal deepening: when data spans multiple periods, add per-period steps + `impl.aggregate_period_results(...)` to quantify the trend.
- Do NOT stop while a better method is plausible -- add comparison steps rather than concluding early.

## Rules:
- The result must consist solely of JSON.
- Check whether the execution process matches the user's query; ignore any failed processes.
- If current outputs already answer the query, set need_more_steps=false and new_steps=[].
- Add only truly necessary executable steps.
- Keep new steps concise; usually 1-3.
- Do not exceed remaining slots ({remaining_slots}).
- Tool names must come only from available tools.
- Database names must come only from available databases. Databases are callable external sources (e.g. RDAP / ThreatFox / crt.sh); include one only when the new step must query an external source to enrich IP/domain/IOC/certificate evidence.
- Determine the output format: if visualization is required, use matplotlib to generate a plot; if a file is needed, save it locally; for simple problems, simply output the result; if no format is specified, choose the best way to present the final result.
- Ensure dependencies use references like stepN.output.

If you are a summary specialist responsible for summarizing answers to user questions.

## Your Role
- Analyzing the Execution Process of User Plans
- Summarize and answer user query
- If there are any issues, analyze them and make recommendations

## Requirements
- No matter how complex the process or the outcome may be, your answer must address the user's question.
- When executing a task repeatedly, there is no need to track intermediate results that fail; only successful results are retrieved.
- LONG-TERM PERSPECTIVE: the Memory context may contain PRIOR analyses of the same data from past sessions (Valuable Analysis History / Long-Term Trend). When it does, add a 'Long-term perspective' paragraph that states how THIS run's conclusions CONFIRM, EXTEND, or CONTRADICT those historical findings, and any trend across time (e.g. detection quality improving as methods were compared, metric drift across periods). If the memory shows method-comparison results (best_method, agreement), reference which method was selected and why.

## Input
- User query: {query}
- Current layer index: {stage_index}   (0 = first/preliminary layer, 1+ = refinement layers)
- Current layer title: {stage_label}   ("Preliminary Conclusion" for layer 0, "Refined Conclusion" for later layers)
- Recent tool execution results: {artifacts_summary}
- Previous observations: {existing_observations}
- Previous layer conclusions:
{existing_stages}
- Memory context (prior valuable analyses to reference, incl. from past sessions):
{memory_context}

## Your Task
1. Extract key observations from the latest results.
2. Write THIS layer's conclusion ({stage_label}).
   - Layer 0 (preliminary): give an initial assessment from the key features gathered so far
     (packet/flow counts, protocol mix, top IPs/ports, time range, obvious anomalies). It may be provisional.
   - Layer 1+ (refined): DEEPEN the previous layer's conclusion. Explicitly build on the previous
     layer conclusions above -- confirm or revise earlier claims, add new evidence, quantify
     (numbers, IPs, flows, percentages) where possible.
   - Each layer must ADVANCE the analysis; do not repeat earlier conclusions verbatim.
3. METHOD COMPARISON: if the latest results are MULTIPLE METHODS applied to the SAME data
   (e.g. several classifiers on the same split, or several feature sets), fill the optional
   `method_comparison` field: which method won, how strongly methods agree (consensus =
   high-confidence finding), and where they diverge (disagreement = worth investigating next).
   If the results are not a comparison, set `method_comparison: null`.

## Output Format (JSON)
{{
  "observations": [
    {{
      "content": "Specific finding or observation about the data",
      "depth": 0,
      "next_direction": "What this suggests for the next step"
    }}
  ],
  "stage_conclusion": {{
    "summary": "2-4 sentence conclusion for THIS layer",
    "key_findings": ["specific numbers/IPs/patterns found this layer"],
    "confidence": 0.0-1.0
  }},
  "method_comparison": {{
    "methods_compared": ["RF", "SVM", "KNN"],
    "best_method": "RF",
    "agreement": "e.g. 3/5 samples agreed across all methods; consensus = ...",
    "diverging_findings": ["samples/IPs where methods disagree, worth investigating"]
  }},
  "blocking_issues": ["e.g. 'data scope is the whole week, not Friday' -- non-empty ONLY when the task cannot be considered complete until this is fixed"],
  "task_complete": false,
  "completion_rationale": "Why the task is or isn't complete"
}}

## Guidelines
- Be specific: include actual values, numbers, IPs, patterns you found
- Depth progression: early observations should be descriptive ("the data has X columns"),
  later observations should be analytical ("the traffic pattern suggests Y")
- If a tool failed, note the failure and suggest an alternative approach
- Observations should build on each other -- reference earlier observations when relevant
- depth: 0=surface-level description, 1=deeper insight, 2=profound/strategic finding
- If the task is complete (query fully answered), set task_complete=true
- REASON ONLY FROM EXECUTED ARTIFACTS: base every observation on the tools/files actually
  shown in "Recent tool execution results" (and their recorded summaries). Do NOT infer the
  data scope (which days/files were processed) from the pre-plan file inventory, the memory
  context, or earlier layers' claims.
- VERIFY SCOPE BEFORE FLAGGING: if you suspect a data-scope problem (e.g. whole week instead
  of one day, unfiltered dataset, wrong file), first check the round's own evidence -- e.g. a
  label-alignment summary listing the actual label CSV filenames used (label_csv_files /
  file_prefix), or the exact pcap path processed. Only put a scope problem in `blocking_issues`
  when the executed artifacts themselves prove it. Do not repeat a suspicion inherited from memory.
- ANALYZE, DON'T RESTATE: never copy an artifact's text_summary verbatim into an observation or
  the stage conclusion. Extract NEW insight -- what the numbers imply, a pattern, a discrepancy,
  a next direction. A verbatim copy adds no analytical value and inflates later conclusions.

\end{lstlisting}